\documentclass[%
 reprint,
 amsmath,amssymb,
 aps,
]{revtex4-2}

\DeclareMathOperator{\Tr}{Tr}
\DeclareMathOperator{\erfc}{erfc}

\usepackage{graphicx}
\usepackage{dcolumn}
\usepackage{bm}

\begin{document}


\title{Interatomic interaction at the aluminum-fullerene $\mathrm{C}_{60}$ interface}

\author{V.V.~Reshetniak}
\email{viktor.reshetnyak84@gmail.com}
\altaffiliation{Troitsk Institute for Innovation and Fusion Research, Russia, 108840, Moscow, Troitsk, st. Pushkovs, 12}

\author{O.B.~Reshetniak}
\affiliation{Troitsk Institute for Innovation and Fusion Research, Russia, 108840, Moscow, Troitsk, st. Pushkovs, 12}

\author{A.V.~Aborkin}
\affiliation{Vladimir State University named after Alexander and Nikolay Stoletovs, Russia, 600000, Vladimir, Gorky st., 87}

\author{A.V.~Filippov}
\affiliation{Troitsk Institute for Innovation and Fusion Research, Russia, 108840, Moscow, Troitsk, st. Pushkovs, 12}
\affiliation{Joint Institute for High Temperatures of the Russian Academy of Sciences, Russia, Moscow, Izhorskaya st., 13}

\date{\today}

\begin{abstract}
We propose a model describing the interatomic interaction at the interface between fullerene $\mathrm{C}_{60}$ and aluminum. Using the density functional theory, we calculate the binding energy and the fullerene's position on the $\mathrm{Al}(111)$ slab. The obtained data are applied to estimate the parameters of the Lennard-Jones potential for carbon and aluminum atoms, which is then used in molecular dynamics simulations. The results of the theoretical study of desorption of fullerenes from an aluminum substrate are in good agreement with those of the experiments from the literature. We also investigate the capillary effects in an aluminum melt with submerged fullerenes. The positive interface surface energy indicates the poor wettability of $\mathrm{C}_{60}$ by the melt. The calculated value of the diffusion relaxation time is approximately two orders of magnitude less than the characteristic coagulation time of fullerenes. The activation character of the coagulation process and the capillary nature of the interaction between fullerenes are discussed. 
\end{abstract}

\maketitle


\section{\label{sec:intro} Introduction}

Carbon particles of various structures are widely used for the development of composite materials with the required properties \cite{chak2020review,hu2006carbon,bakshi2010carbon,kumar2014graphene}. In particular, aluminum matrix composites filled with $\mathrm{C}_{60}$ fullerenes are of considerable interest \cite{khalid2003study,korobov2016mechanical,shin2015mechanical}.

One of the key issues in creating composite materials is to ensure the interaction between the matrix and the filler. The interaction of $\mathrm{C}_{60}$ fullerenes with different crystalline aluminum surfaces has been studied experimentally \cite{maxwell1995c,maxwell1998electronic,johansson1996scanning,johansson1998adsorption,fasel1996orientation,hamza1994reaction} and theoretically \cite{stengel2003adatom}. It is known from Refs~\cite{maxwell1995c, maxwell1998electronic, johansson1996scanning, johansson1998adsorption} that when $\mathrm{C}_{60}$ comes into contact with aluminum surfaces, the covalent chemical bond appears between $\mathrm{Al}$ and $\mathrm{C}$ atoms. Using the data from scanning tunneling microscopy (STM), photoelectron spectroscopy (PES), and X-ray absorption spectroscopy (XAS), two different surface reconstruction models were proposed: $2\sqrt{3} \times 2\sqrt{3}$ and $6 \times 6$. It was established that the binding energies differ for different reconstruction types. The structure of the $2\sqrt{3} \times 2\sqrt{3}$ fullerene layer is stable at lower temperatures, and at 490~K the coating acquires a $6 \times 6$ structure. Desorption of the remaining fullerenes from the surface appears at 730~K.

According to the microscopy data by Maxwell et al. \cite{maxwell1995c} in the $2\sqrt{3} \times 2\sqrt{3}$ reconstruction, there are two different ways of arranging fullerenes on the $\mathrm{Al}(111)$ surface. They are characterised by different binding energies and distances between the molecule and the substrate. The model by Stengel et al. \cite{stengel2003adatom} explains this effect by the displacement of the molecule towards the substrate with a vacancy. Subsequent studies performed for various metals' surfaces with the dense packing of atoms indicate that fullerenes tend to be located above the vacancies on these substrates. The crystal structure of the metal determines the type of the reconstruction \cite{shi2012adsorbate}.

The interaction between fullerenes and the $\mathrm{Al}(100)$ and $\mathrm{Al}(110)$ surfaces is rarely studied. Maxwell et al. \cite{maxwell1998electronic} showed that when $\mathrm{C}_{60}$ interacts with $\mathrm{Al}(110)$, carbon and aluminum atoms form covalent bonds, and fullerenes desorb at 730~K. However, they did not observe any different ways of arranging the molecules on these surfaces and the corresponding desorption peaks.

We applied the Lennard-Jones potential to describe the interaction between $\mathrm{Al}$ and $\mathrm{C}$ atoms for $\mathrm{Al} - \mathrm{C}_{60}$ systems

\begin{equation}\label{eq1}
    u(r) = 4\varepsilon \left[ \left( \frac{\sigma}{r} \right)^{12} - \left( \frac{\sigma}{r} \right)^{6} \right],
\end{equation}

\noindent where $\varepsilon$ is the depth of the potential well and $\sigma$ is the point where the potential energy is zero. To assess these parameters, we performed the \textit{ab initio} calculation of the fullerene-$\mathrm{Al}(111)$ slab interaction and then analyzed the results with an analytical model based on a homogeneous medium approach. A detailed discussion of the proposed model can be found in Section~\ref{sec:interact}.

The analytical model of temperature-programmed desorption (TPD) based on the approximations of the transition state theory was applied for comparison with the experiment. We estimated the unknown parameters of the model from the preliminary \textit{ab initio} calculated values of the distance and the binding energy of the $\mathrm{C}_{60}$ molecule with the $\mathrm{Al}(111)$ surface.

In the TPD method, the system's temperature increases according to a certain law (usually linear) in a given range of values. During the heating, the flux of molecules from the surface is monitored. The advantage of the approach is the possibility of a detailed analysis of desorption, which allows determining the binding energies with high accuracy for the molecules on different sites on the substrate. A detailed description of the method can be found, for example, in review \cite{king1975thermal}. Hamza et al. \cite{hamza1994reaction} applied the TPD to investigate the interaction between $\mathrm{C}_{60}$ fullerenes and crystalline aluminum surfaces. They determined the dependence of the coverage density $n$ on temperature $T$ of a monolayer of fullerenes at a constant heating rate of 3~K/s.

This paper proposes a multiscale study of the desorption kinetics in different temperature ranges. We calculated the dependence $n(T)$ both analytically and numerically, using molecular dynamics (MD) simulations, for time intervals of about 1~ns. To speed up the desorption process, we used an underestimated values of the binding energies. The analytical model and the binding energy from our DFT calculations were applied for laboratory time intervals comparable to the characteristic times from the experiments of Hamza et al. \cite{hamza1994reaction}. The comparison suggests that the proposed analytical model and the interatomic potential reproduce the known experimental data with high accuracy. The data from the MD simulations are in reasonably good agreement with the analytical ones. The obtained results are discussed in Section~\ref{sec:tpd}.

The authors of \cite{altman1993interaction, li2009surface, modesti1995stable, villagomez2020first} indicate that the values of the binding energies of fullerenes with the substrate, the structure and the desorption temperatures are very close for the surfaces of some different metals. Particularly, the proposed results for the aluminum substrate are very close to those obtained for noble metal substrates, which allows us to hope that the results obtained for $\mathrm{Al}-\mathrm{C}_{60}$ can be generalised for $\mathrm{Au}-\mathrm{C}_{60}$ and $\mathrm{Ag}-\mathrm{C}_{60}$ systems without significant refinement. However, the possibility of such generalisation requires additional analysis, which is beyond the scope of this study.

In this paper, we also consider capillary phenomena at the interface between molten aluminum and submerged fullerenes. We calculated the coagulation rate, and the results indicate the presence of long-range interaction between fullerenes. It is found that the processes of coagulation and diffusion of fullerenes have different characteristic relaxation times and can be separated into ``slow'' and ``fast''. We estimated the Laplace pressure $p$ and the surface tension $\gamma$ of the interface from the MD simulations. Then we calculated the bulk modulus from the change in the volume of fullerenes due to the Laplace pressure. The results of the study of capillary phenomena are discussed in Section~\ref{sec:capillary}.

\section{\label{sec:model}Description of models and methods of the numerical research}
\subsection{\label{sec:dft}\textit{Ab initio} calculation of the fullerene-$\mathrm{Al}(111)$ slab interaction}

DFT calculations were performed using the CP2K program \cite {hutter2014cp2k, vandevondele2005quickstep}. We estimated the energies and forces, considering all the electrons in the framework of the Gaussian and augmented-plane-waves (GAPW) method \cite{lippert1999gaussian} using the pob-TZVP basis set from \cite{peintinger2013consistent}. Exchange and correlation of electrons were considered with the Purdue~-- Burke~-- Ernzerhof (PBE) model \cite{perdew1996generalized}.

First, we optimized the atomic positions and the parameters of the face-centered cubic (fcc) crystal cell of the aluminum using DFT. The calculated value of the lattice parameter $a = 4.0775$~\AA\ within 1\% coincided with the experimental one, 4.0469~\AA, taken from Wyckoff's book \cite{wyckoff_crystal_1963}. Then we used the obtained data to construct an $\mathrm{Al}(111)$ slab with the orthorhombic supercell of $5 \times 6$ in size and four atomic layers thick. We placed the slab in the $XY$ plane and set the distances between the surface atoms and the simulation cell boundaries in the $Z$ direction of 10~\AA. The integrals over the Brillouin zone were calculated using a $3 \times 3 \times 1$ Monkhorst-Pack grid \cite{monkhorst1976special}. We estimated the surface energy $E_{s}$ and the vacancy formation energy $E_v$ for the preliminary verification of the model. The value of $E_{s} = 1.06$~eV from our simulations is in good agreement with the results of \textit{ab initio} calculations from Refs~\cite{schochlin1995structure,jacobs2002,fiolhais2003extraction}, where values in the range from 0.87~eV to 1.46~eV were obtained, and with the experimental data from paper \cite{tyson1977surface}, $E_{s} = 1.14-1.18$~eV. The calculated vacancy formation energy $E_{v} = 0.65$~eV agrees with the data reported by Kiejna \cite{kiejna2003vacancy}, where the value of 0.61~eV was obtained.

\begin{figure}
\begin{minipage}{0.49 \linewidth}
\includegraphics[width=1\linewidth]{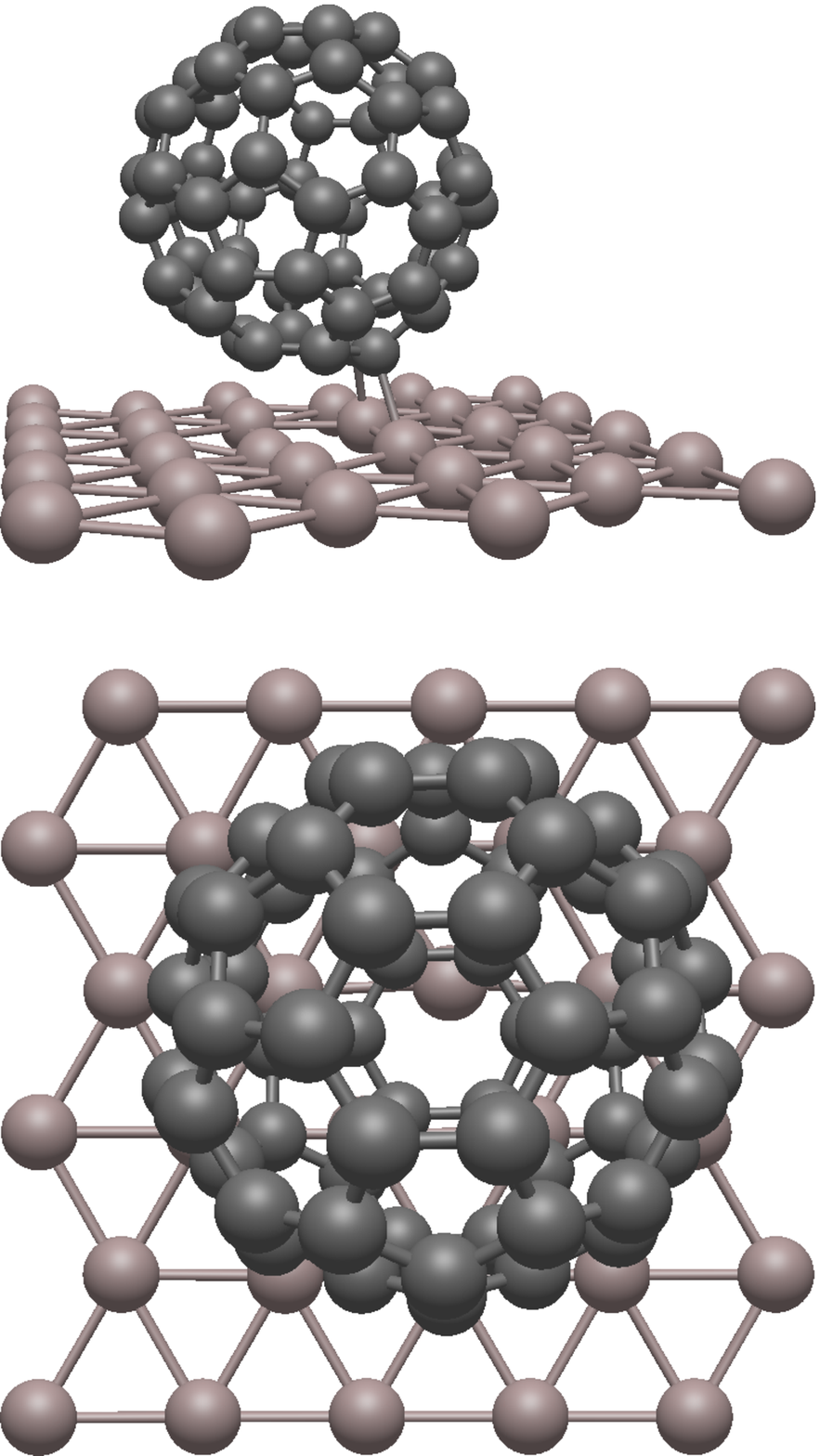} a%
\end{minipage}
\begin{minipage}{0.49 \linewidth}
\includegraphics[width=1\linewidth]{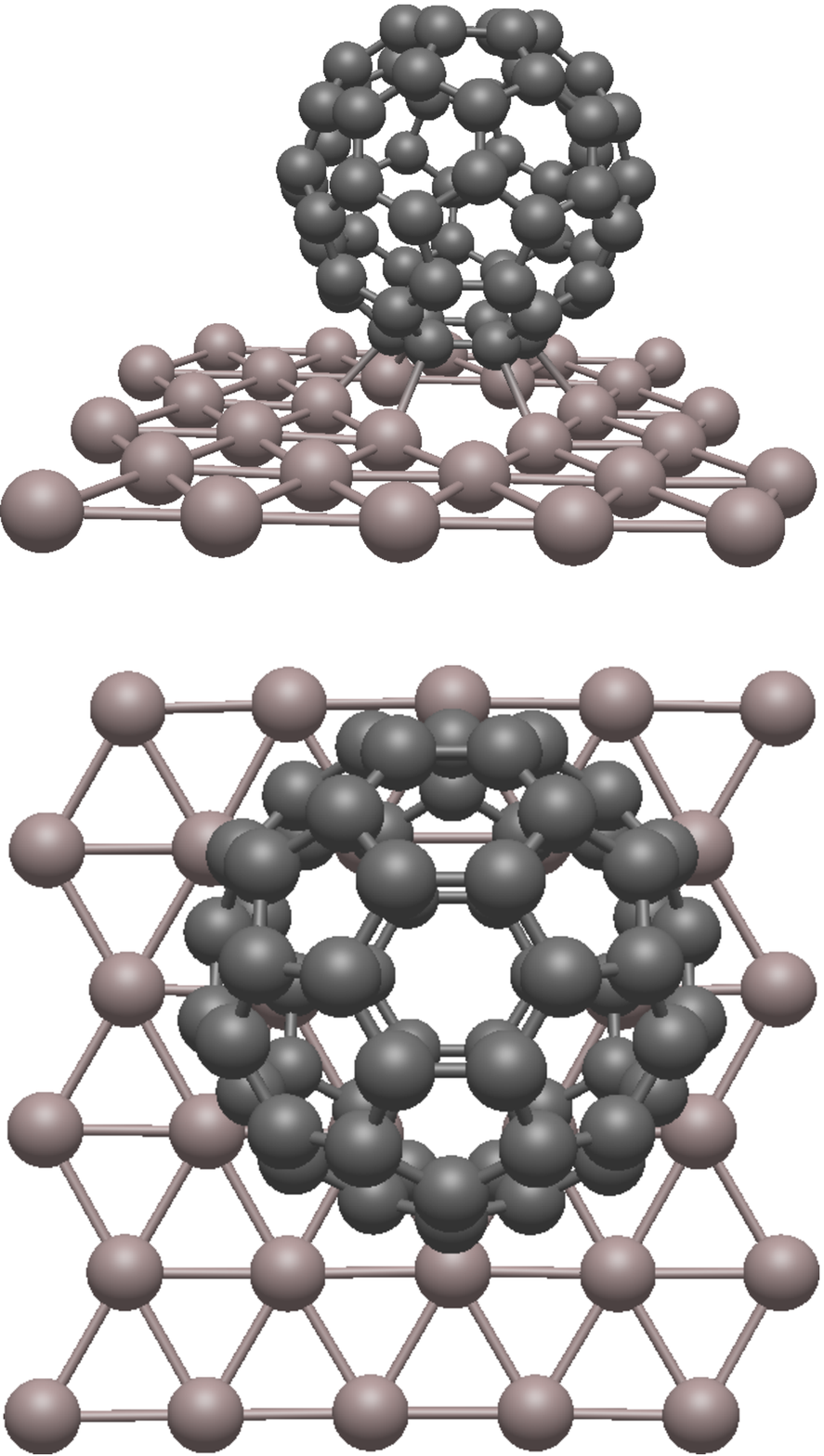} b%
\end{minipage}

 \caption{\label{fig:fig1}%
  Scheme of the arrangement of atoms in the interaction of fullerene $\mathrm{C}_{60}$ with the $\mathrm{Al}(111)$ surface. 
  (a)~-- the hollow-6 configuration, substrate without a vacancy;
  (b)~-- the fullerene is located above the vacancy on the substrate.
 }%
\end{figure}

We considered two variants of the fullerene positioning on the aluminum slab. In the first case, we placed $\mathrm{C}_{60}$ above one of the aluminum atoms, according to the hollow-6 model from Ref~\cite{stengel2003adatom}. If there is no vacancy on the substrate, this position corresponds to the system's most stable state. In  the  second  case  the  fullerene  was placed  over  the  vacancy  on the  aluminum surface. The relative positions of the $\mathrm{C}_{60}$ molecule and aluminum atoms are shown schematically in Fig.~\ref{fig:fig1}.

\subsection{\label{sec:empirical}Atomistic simulations using the empirical force fields}

We used the LAMMPS program \cite{plimpton1995fast} to perform all calculations with the empirical interatomic potentials. The Tersoff potential \cite{tersoff1989modeling,tersoff1988} was applied to consider the interaction between carbon atoms. The embedded atom model (EAM) with the parametrization proposed by Sheng et al. \cite{sheng2011highly} was employed for aluminum atoms. We used the Lennard-Jones potential of form \eqref{eq1} for the interaction between carbon and aluminum atoms. Estimating the parameters of the potential is discussed in Section~\ref{sec:interact}. Equations of motion were integrated numerically with the timestep of $d t = 0.2$~fs.

We simulated the fullerenes desorption from the solid $\mathrm{Al}(111)$ slab and the molten aluminum film. The film with a thickness of 10 atomic layers was placed parallel to the $XY$ plane. The size of the film corresponded to the supercell $36 \times 36$ for the unit cell of $\mathrm{Al}(111)$. We applied the periodic boundary conditions at the computational domain boundaries along the $X$ and $Y$ axes. At the upper edge $z = z_{\max}$, located at a distance of 10~nm above the upper atomic layer, a ``reflecting wall'' boundary condition was used. At the lower boundary, we applied the Lennard-Jones 9-3 potential

\begin{equation}\label{eq2}
    w = \frac{2}{3} \pi n_{1} \varepsilon \sigma^{3} \left[ \frac{2}{15} \left( \frac{\sigma}{z} \right)^{9} - \left(\frac{\sigma}{z} \right)^{3} \right],
\end{equation}

\noindent where $n_{1}$ is the average density of the aluminum atoms in the substrate. The following values from the paper by Heinz et al. \cite{heinz2008accurate} were taken as parameters for aluminum: $\varepsilon = 0.174$~eV, $\sigma = 2.925$~\AA. 

One can obtain the potential in form \eqref{eq2} by estimating the interaction energy between an atom and a semi-infinite homogeneous substrate and assuming that the interatomic interactions have the Lennard-Jones type \cite{smith1979lennard}. To do this, one needs to calculate the following integral over the volume of the substrate

\begin{equation}\label{eq3}
    w = 2 \pi n_{1} \int_{H}^{\infty} u(L) L (L - H) d L.
\end{equation}

\begin{figure*}
\begin{minipage}{0.32 \linewidth}
\includegraphics[width=1\linewidth]{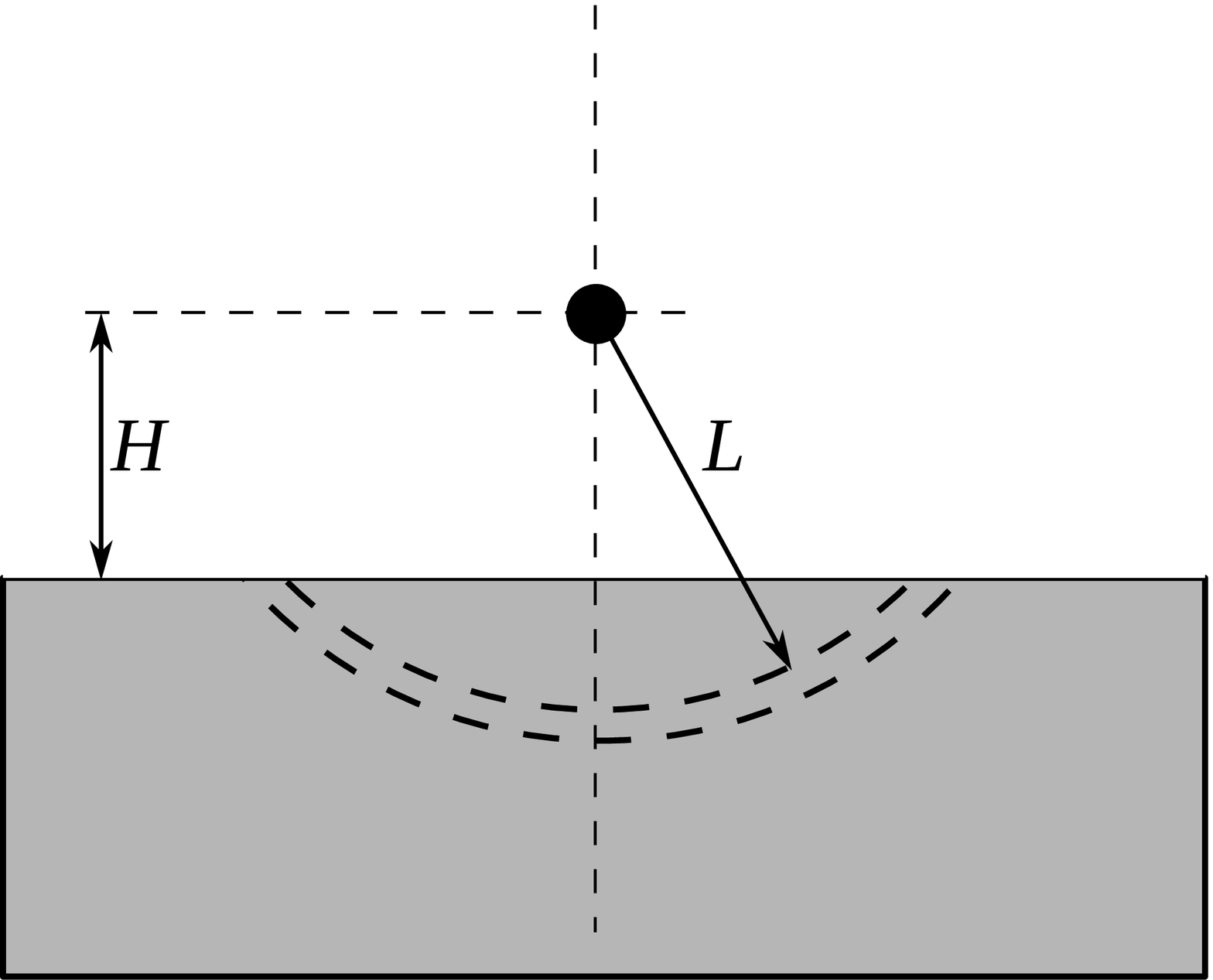}%
\vspace{0.5cm} 
\end{minipage}
\begin{minipage}{0.32 \linewidth}
\includegraphics[width=1\linewidth]{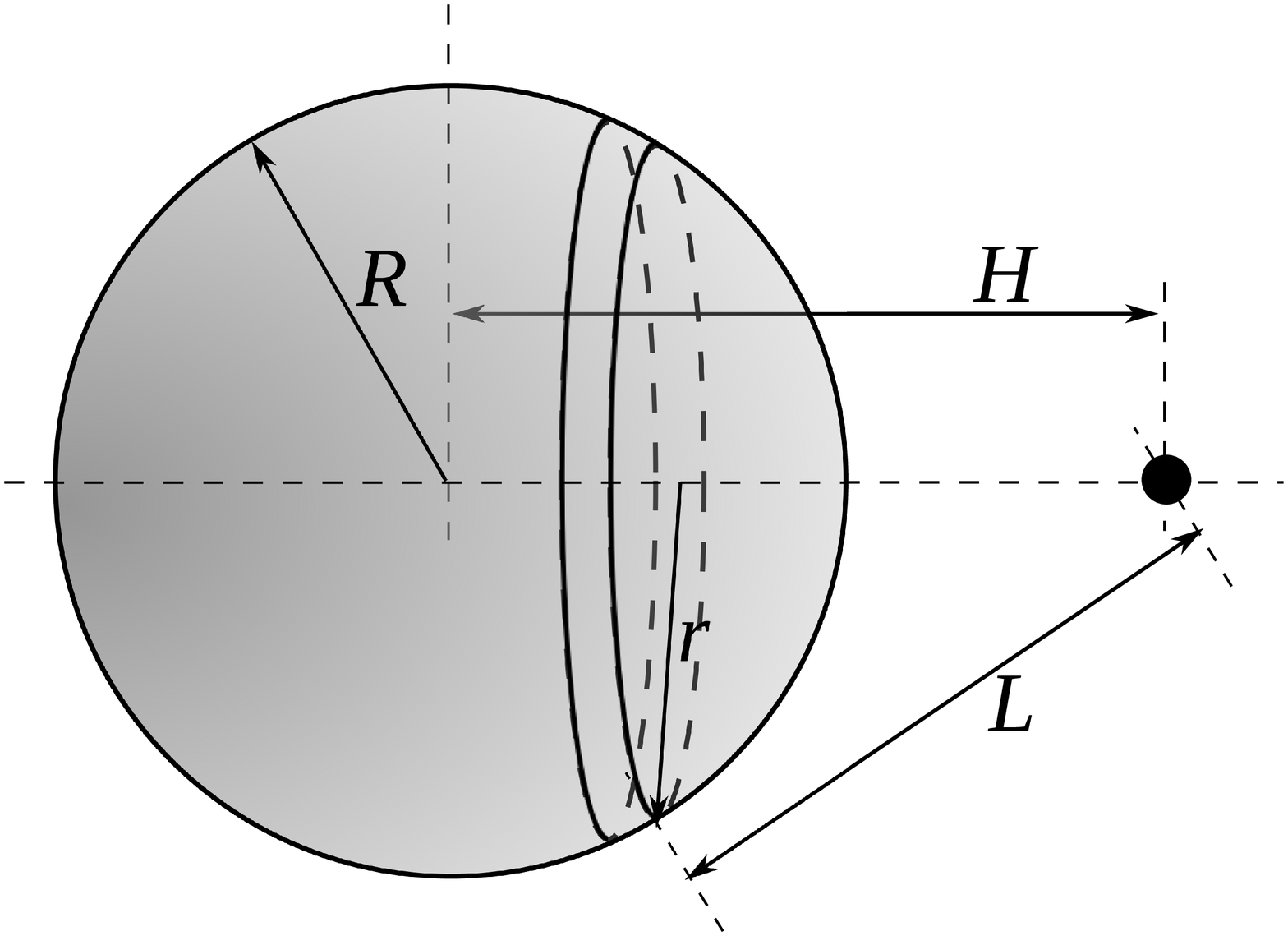}%
\vspace{0.5cm} 
\end{minipage}
\begin{minipage}{0.32 \linewidth}
\includegraphics[width=1\linewidth]{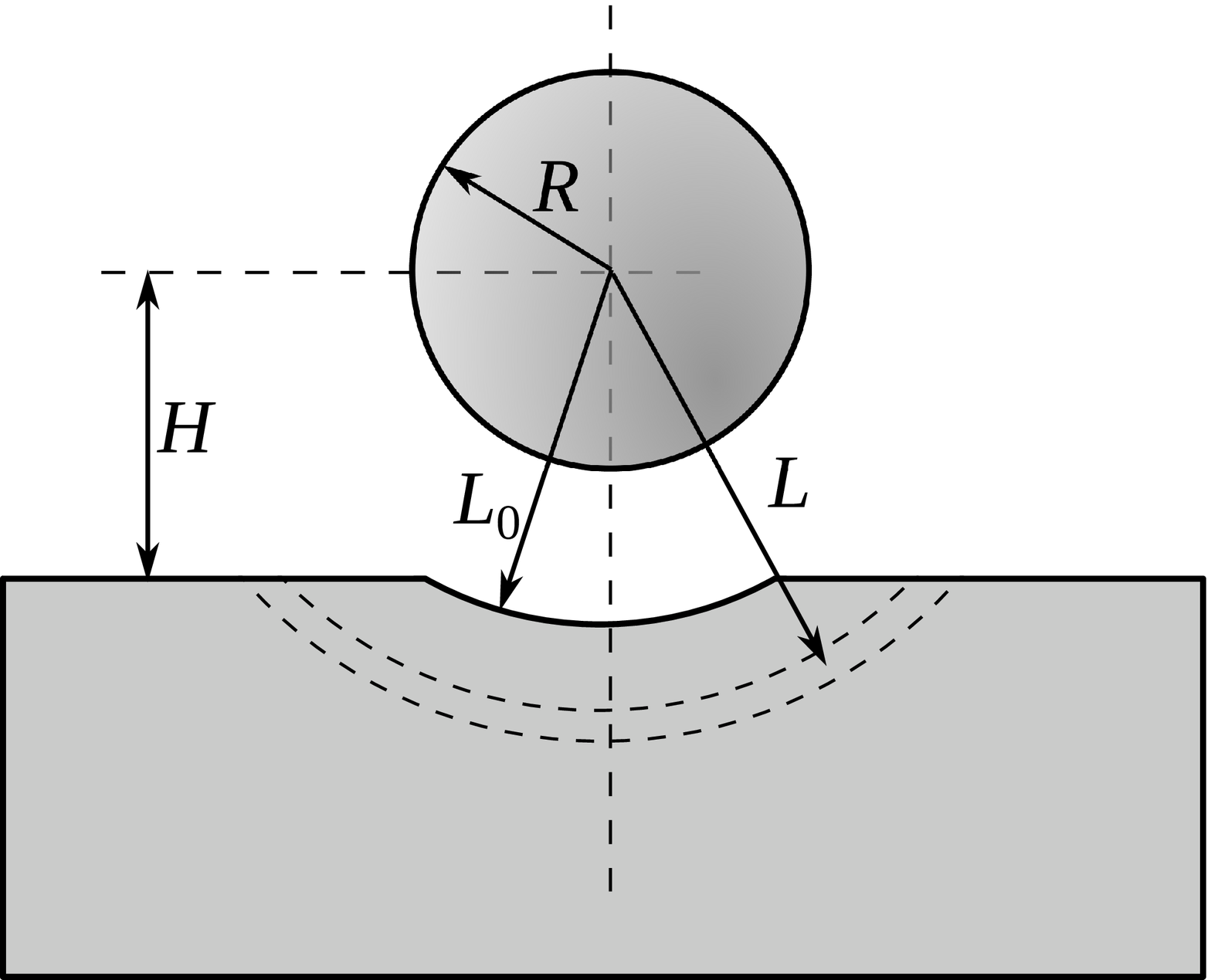}%
\vspace{0.5cm} 
\end{minipage}
\hspace{0.3 \linewidth} a \hspace{0.3 \linewidth} b \hspace{0.3 \linewidth} c

 \caption{\label{fig:fig2}%
  Estimation of the interaction between particles using the approximation of a homogeneous substance with an average density. 
  (a)~-- decomposition of the substrate by spheres of radius $L$ for integration of \eqref{eq3};
  (b)~-- the splitting of the sphere by parallel circles for estimation of the interaction between an atom and a homogeneous sphere;
  (c)~-- decomposition of the substrate by spheres for calculation of the interaction between a sphere and a substrate with a vacancy.
 }%
\end{figure*}

\noindent Here $L$ is the distance from the atom to the points of the volume element $d V(L)$ obtained by decomposition the substrate by spherical surfaces of radii $L$ and $L + d L$, as shown in Fig.~\ref{fig:fig2}a.

Simulation of the TPD was carried out with the surfaces of crystalline and liquid aluminum. We preliminary stabilized the film at a given initial temperature 400~K in the first case and 1000~K in the second one. Then 108 fullerenes were placed on the surface according to the reconstruction model $2\sqrt{3} \times 2\sqrt{3}$, and we stabilized the system for 20~ps. After stabilisation, we heated the system to the target temperature, 800~K for the crystal and 2200~K for the liquid. The heating duration was 1~ns. We used a Langevin thermostat with the temperature relaxation parameter set to 20~fs to control the temperature. To achieve desorption at the given time scales and temperature range, we reduced the value of the coupling parameter $\varepsilon$ of the potential \eqref{eq1} for carbon and aluminum atoms. We performed the simulations for the crystal substrate with the value $\varepsilon_{1}$, four times lower than the initial $\varepsilon$. For the desorption from the melt surface the value $\varepsilon_{2} = 0.7 \varepsilon$ was set. In total, we performed five statistically independent calculations and estimated the averaged temperature dependence of the coverage density $n(T)$.

We considered a system of 174960 atoms of aluminum and carbon to calculate the vibrational spectral density of the centers of mass of the molecules. The system included an $\mathrm{Al}(111)$ slab of $108 \times 108 \times 10$ atoms and 972 fullerenes. In this case we probed only the crystalline slab $\mathrm{Al}(111)$ and performed calculations only at $T=300$~K without decreasing the parameter $\varepsilon$. We estimated the velocities of the centers of mass of the molecules at different time points during the calculations. Then we used the results to compute the spectral density of the oscillations $S(\omega)$ according to the formula \cite{landau1980statistical}:

\begin{equation}\label{eq4}\nonumber
    S = 2 \int_{0}^{\infty} \langle v_{z}(t) v_{z} (0) \rangle \cos(\omega t) d t.
\end{equation}

Additionally, we calculated the surface tension at the $\mathrm{Al} - \mathrm{C}_{60}$ interface at a temperature of 1000~K for the aluminum melt and fullerenes immersed in it. A system of 61500 atoms was considered: 54000 aluminum atoms and 7500 carbon atoms (125 fullerenes). This ratio of components approximately corresponds to the fullerene concentration of 5.8 wt\%. Initially, we located the aluminum atoms in the nodes of the fcc lattice and the fullerenes in the nodes of the simple cubic lattice. The periodic boundary conditions were applied for all boundaries of the computational domain. Then we stabilized the system at 1000~K and zero pressure for 100~ps, using the algorithm proposed by Shinoda et al. \cite{shinoda2004rapid}. After the stabilisation, we turned off the thermostat and the barostat and performed all subsequent calculations in a microcanonical ensemble. The simulation time interval was 10~ns, and we analyzed the parameters only for the data obtained in the NVE ensemble.

In our simulation, the temperature, the potential energy and the local pressure per carbon atom were estimated. To calculate the local pressure, we used the algorithm from paper \cite{thompson2009general} and the expression:

\begin{equation}\label{eq5}
    p = \frac{n_{V}}{3} \Tr(m v_{a} v_{b} + W_{a b}),
\end{equation}

\noindent where $m$ is the atom's mass, $v$ is its velocity, the subscripts $a$ and $b$ correspond to the vectors' components, and $W_{ab}$ is the virial and $\Tr$ is the trace of a matrix. To calculate the density $n_{V}$ of carbon atoms, we assumed that the fullerene boundary coincides with a thin spherical interface, the radius of which was set approximately as $ R_{f} \approx R + 2^{-5/6} \sigma$. We estimated the value of $R$ from the MD data as the averaged distance between the carbon atoms and the fullerene center of mass. We averaged the results over the ensemble and over the time interval of 1~ns. The surface tension was calculated using the Laplace pressure

\begin{equation}\label{eq6}
    \gamma = \frac{p R_{f}}{2}.
\end{equation}

Then we investigated the temperature dependence of the fullerene radius $R(T)$ for the free molecules at a zero pressure. We used averaging over an ensemble of 1000 fullerenes and a time of 10000 steps (2~ps). The temperature range from 100 to 2000~K was considered. Then, the coefficient of thermal expansion (CTE) was estimated as:

\begin{equation}\label{eq7}
    \beta = \frac{1}{V_{f}} \left( \frac{\partial V_{f}}{\partial T} \right)_{p=0}.
\end{equation}

The subscript $f$ in expression \eqref{eq7} and below means that in estimating the volume $V_{f}$ we use the value of the interface radius $R_{f}$, defined above. 

When a fullerene is immersed in the melt, the volume changes by the value $\Delta V_{f}$ under the action of the Laplace pressure $p$. Knowing these values from the MD, we calculated the bulk modulus of a fullerene by the formula:

\begin{equation}\label{eq8}
    K_{T} = -V_{f} \left( \frac{\partial p}{\partial V_{f}} \right)_{T} \approx -V_{f} \frac{p}{\Delta V_{f}}
\end{equation}

It is necessary to emphasize some uncertainty in calculating the volume, density, and surface area of fullerene caused by the smallness of the molecule, whose radius is commensurate with the interatomic distances. Due to the smallness of the fullerene radius, the physical meaning of the macroscopic characteristics, such as bulk modulus or surface tension, for individual molecules requires clarification. Since the fluctuations of microscopic quantities for small molecules are significant, the parameters need to be averaged over the ensemble. In composite materials, such averaging is easy to perform due to the presence of many particles in actual samples. Besides, the values of the volume- or density-related thermodynamic parameters $\beta$, $\gamma$ or $K$ are sensitive to the way of calculating the particle's radius. One may perform the estimations with or without account for the finite volume of the carbon atom. Without specifying the method for calculating the fullerene radius, these values do not make sense. 

Khabibrakhmanov and Sorokin \cite{khabibrakhmanov_2020} discussed in detail the influence of the method for calculating the bulk modulus of carbon nanoparticles, including fullerenes, on the result obtained. They showed that the application of a model based on the use of the $\mathrm{C} - \mathrm{C}$ chemical bond for calculating the bulk stiffness modulus eliminates the ambiguity associated with the choice of the nanoparticle volume. In this case, nanoparticles are compressed by the method of coordinate renormalization. Note that in the condition of particle compressibility under the action of the external environment, which usually occurs in experiments, this problem arises again and manifests itself in the ambiguity of the elastic properties of the compressing substance. This effect is due to the application of continuum models to microscopic objects and can be clearly illustrated by comparing the bulk moduli of the face centered cubic lattice of fullerite \cite{kobelev_elastic_1997}, where the intermolecular interaction of fullerenes occurs according to the dispersion mechanism and the bulk modulus is 10.8~GPa, and superhard fullerite \cite{kvashnina_fullerite-based_2017,perottoni_first-principles_2002}, where fullerenes are covalently bound and the calculated values of the bulk modulus are in the range from 236~GPa to 304~GPa. In the first case, the elastic compression of a substance is accompanied by a predominant change in the distances between neighboring molecules, while in the second case, a change in the intramolecular interatomic distances is also significant. We consider it possible to study the elastic properties of fullerenes in these systems by considering the ensemble-averaged deformations of ``individual'' molecules under the action of the external environment. When analyzing the pressure dependence of the radii of molecules, the calculation of the bulk moduli of fullerenes with different compressing substances would give significantly smaller differences than those obtained for crystals. However, the density and elastic moduli of the compressing substance, which in this case will occupy an additional volume per space between the individual fullerenes, will be ambiguously determined. This ambiguity can be eliminated by including in the model the dependence of the volumes of the compressible and compressing substances on the interaction potential between atoms belonging to different molecules.

In the present work, we propose to approximately take into account this ambiguity by calculating the radius $R_{f}$ of a thin spherical interface between aluminum and fullerene, with $R_{f}$ being determined by the parameters of the interaction potential of $\mathrm{Al} - \mathrm{C}$ atoms. Assuming the value of $R_{f}$ to be the inclusion radius and using averaging over the ensemble, it is possible to obtain thermodynamic values well defined for the $\mathrm{Al} - \mathrm{C}_{60}$ system both for the inclusion and for the matrix. Note that in this work, the value of $R_{f}$ was calculated in the zeroth approximation, taking into account the dependence of $R_{f}$ on the position of the minimum, but without allowance for the potential stiffness \eqref{eq1}.

\section{\label{sec:res} Results and discussion}
\subsection{\label{sec:interact}Interaction between fullerenes and the substrate}

We used the following simplifying assumptions for the theoretical analysis of the numerical results of modelling the interaction of fullerenes with an aluminum substrate. We assumed that aluminum atoms are distributed over the substrate of infinite width and thickness uniformly, with a given average density $n_{1}$. The value of $n_{1} \approx 0.0590$~\AA$^{-3}$ was estimated with the unit cell optimization within the DFT framework. The fullerene was also considered as a thin homogeneous sphere with an average density of the surface distribution of atoms, $n_{2} = N / (4 \pi R^{2}) \approx 0.3761$~\AA$^{-2}$. We calculated the sphere's radius as the average distance between the carbon atoms and the fullerene's center of mass after the relaxation of the atomic positions. The bond lengths were 1.41~\AA\ between the atoms located at the vertices of the pentagonal structural elements of fullerene and 1.45~\AA\ between the atoms in the hexagons' vertices. This result is in good agreement with the experimental data, 1.40~\AA\ and 1.46~\AA\ (see ref~\cite{dresselhaus_science_1996} and references therein).

These assumptions do not allow us to consider the dependence of the binding energy on the position of the fullerene on the aluminum surface, which is implicitly assumed to be weak. The validity of the accepted hypothesis is not apparent, but the following experimentally known facts support it. First, according to Maxwell et al. \cite{maxwell1998electronic, johansson1998adsorption}, fullerenes are highly mobile on an aluminum substrate even at room temperature. Second, the temperature dependence of the coverage density $n(T)$ from the TPD experiments \cite{hamza1994reaction} can be accurately approximated by an analytical function with a single inflection point. Therefore, we assume that the changes in the binding energy due to the in-plane displacement of fullerenes do not play a significant role in the experiments.

Using the uniform medium approximation and the interatomic interaction potential \eqref{eq1} allows us to obtain formula \eqref{eq2} for the interaction between an atom and a substrate. Similarly, by integrating over the surface of the sphere (see Fig.~\ref{fig:fig2}b), we obtain the expression for the energy of interaction between an atom and a fullerene:

\begin{equation}\label{eq9}
    \epsilon = 8 \pi \varepsilon \sigma^{2} n_{2} \frac{R}{H} \left[\frac{1}{10}(\zeta_{10}^{-} - \zeta_{10}^{+}) - \frac{1}{4}(\zeta_{4}^{-} - \zeta_{4}^{+})\right],
\end{equation}

\noindent where the notation $\zeta_{m}^{\pm} = [\sigma / (H \pm R)]^{m}$ is used to shorten the entry, $H$ is the distance from the atom to the center of the sphere, $R$ is the radius and $m$ is the exponent.

Suppose that the vacancy has the shape of a spherical segment formed by the intersection of the aluminum substrate and the sphere centered in the fullerene's center of mass (see Fig.~\ref{fig:fig2}c). We evaluated the parameter $L_{0}$ using equality between the volumes of the cut segment and the aluminum atom and assuming that the location of the fullerene above the substrate corresponds to the minimum potential energy, $H = H_{0}$,

\begin{equation}\label{eq10}\nonumber
    n_{1}^{-1} = \pi \left[ \frac{2}{3} (L_{0}^{3} - H_{0}^{3}) - H_{0}(L_{0}^{2} - H_{0}^{2}) \right].
\end{equation}

Integrating \eqref{eq9} over the volume of aluminum, we obtain the dependence of the potential energy on the distance $H$:

\begin{eqnarray}\label{eq11}
    W(H, L_{0}) = &&\frac{2}{3} \pi^{2} \varepsilon R n_{1} n_{2} \sigma^{3} \nonumber
    \\
    &&\times \left[ \frac{1}{30} (I_{9}^{-} - I_{9}^{+}) - (I_{3}^{-} - I_{3}^{+}) \right],
\end{eqnarray}

\noindent where we introduced the notation 

\begin{equation}\nonumber
    I_{m}^{\pm} = [m L_{0} - (m-1) H \pm R] \zeta_{m}^{\pm},
\end{equation}
 
\noindent and $m$ takes the values 3 or 9.

Without a vacancy, $L_{0} = H$, and formula \eqref{eq11} reduces to the form:

\begin{eqnarray}\label{eq12}
    W(H) = &&\frac{2}{3} \pi^{2} \varepsilon R n_{1} n_{2} \sigma^{4} \nonumber
    \\
    && \times \left[ \frac{1}{30} (\zeta_{8}^{-} - \zeta_{8}^{+}) - (\zeta_{2}^{-} - \zeta_{2}^{+}) \right].
\end{eqnarray}

The values of $H = H_{0}$ and $W(H_{0}) = W_{0}$, estimated from DFT, correspond to the minimum of the potential energy. Applying the minimum condition to expression \eqref{eq12} and equating the resulting values $H_{0}$ and $W_{0}$ to the values previously calculated \textit{ab initio}, we write a system of equations for determining the potential parameters \eqref{eq1}

\begin{eqnarray}\label{eq13}
\sigma = &&(H_{0}^{2} - R^{2})\left[ \frac{15}{2} \frac{(H_{0} + R)^{3} - (H_{0} - R)^3}{(H_{0} + R)^{9} - (H_{0} - R)^9} \right]^{1/6},
    \\
\label{eq14}
\varepsilon = &&\frac{3 W_{0}}{2 \pi^{2} n_{1} n_{2} R \sigma^{4}} \left[\frac{1}{30}(\zeta_{8}^{-} - \zeta_{8}^{+}) - (\zeta_{2}^{-} - \zeta_{2}^{+}) \right]^{-1}.
\end{eqnarray}

We calculated the binding energy  $W_{0}$ of the fullerene to the substrate according to the DFT data as

\begin{equation}\label{eq15}
    W_{0} = E_{\mathrm{Al}-\mathrm{C}} - (E_{\mathrm{Al}} + E_{\mathrm{C}}),
\end{equation}

\noindent where $E_{\mathrm{C}}$ and $E_{\mathrm{Al}}$ are the potential energies of the free molecule and the aluminum substrate, respectively; and $E_{\mathrm{Al}-\mathrm{C}}$ is the energy of the fullerene interacting with the substrate.

The DFT calculations for fullerene located on an ideal $\mathrm{Al}(111)$ surface give the value of the binding energy $W_{0} = -0.98$~eV and the distance $H = 5.50$~\AA. If there is a vacancy on the surface, the parameters differ significantly: $W_{0} = -2.04$~eV and $H = 4.99$~\AA. This effect corresponds to the one described by Stengel et al. \cite{stengel2003adatom}. As one of the probable reasons, Stengel et al. \cite{stengel2003adatom} noted a strong steric effect observed during the relaxation of the atomic positions. The aluminum atom located closest to the fullerene was shifted deeper relative to the substrate surface during the optimization. The excessive overlap of the aluminum and carbon atoms caused this shift and increased the distance $H$ and the energy $E_{\mathrm{Al}-\mathrm{C}}$. It is evident for the reasons of geometry that removing the specified aluminum atom will increase the bond strength between the fullerene and the substrate and shift the molecule toward the surface. Furthermore, the effect of strengthening the bond can be caused by an increase in the chemical activity of aluminum atoms due to a decrease in their coordination number.

The use of the empirical model of interatomic interaction \eqref{eq1} allows us to separate the possible causes of this effect. On the one hand, since the potential parameters do not depend on the relative position of the aluminum atoms, the model is not sensitive to the possible change in the chemical activity of the atoms due to a vacancy. On the other hand, the potential \eqref{eq1} allows for the repulsion of atoms at short distances and makes it possible to reproduce the interaction features associated with the steric effect. Additionally, expression \eqref{eq11} enable us to consider the change in the interaction energy and the displacement of the fullerene. We obtained expression \eqref{eq11} assuming a specific shape of the vacancy with the center of curvature coinciding with the center of the fullerene (see Fig.~\ref{fig:fig2}c). This assumption simplifies the integration procedure significantly. One can assume that using a more realistic model would more accurately reproduce the results of \textit{ab initio} calculations. Due to the specific shape of the vacancy, consistent with the shape of the fullerene surface, the analytical evaluation with formula \eqref{eq11} may overestimate the absolute value of the binding energy and underestimate the equilibrium distance $H_{0}$. Nevertheless, we believe that the proposed simplified model can be helpful to analyze the discussed effects, if supplemented with more rigorous numeric calculations.

We estimated the interaction energy $W_{0}$ at zero temperature using both the \textit{ab initio} model and the empirical potential of form \eqref{eq1} with the parameters defined by equations \eqref{eq13} and \eqref{eq14}. Our results and the results obtained by Stengel et al. \cite{stengel2003adatom} are listed in Table~\ref{tab:tab1}. For the convenience of comparison Table~\ref{tab:tab1} is supplemented with the averaged values of the $\mathrm{C}-\mathrm{Al}$ bond lengths. Optimization of atomic positions and calculation of energy using the potential \eqref{eq1} in the LAMMPS program makes it possible to consider the actual location of the atoms and does not use any assumptions related to the homogeneity of the substance. Comparing the results of the empirical calculations with \textit{ab initio} calculations allows us to judge the applicability of the homogeneous substance model and expressions \eqref{eq13} and \eqref{eq14} to parameterize the potential \eqref{eq1}.

\begin{table*}
\caption{\label{tab:tab1}Interaction of fullerene with the $\mathrm{Al}(111)$ surface at zero temperature, calculated using the DFT and the potential \eqref{eq1}}.
\begin{ruledtabular}
\begin{tabular}{lcccccc}
 &$W_{0}$,~eV&$H_{0}$,~\AA&$\mathrm{C}-\mathrm{Al}$,~\AA&$W_{0}^{vac}$,~eV&$H_{0}^{vac}$,~\AA&$\mathrm{C}-\mathrm{Al}^{vac}$,~\AA\\ \hline
 DFT (this work)& -0.98 & 5.50 & 2.24 & -2.04 & 4.99 & 2.19 \\
 DFT \cite{stengel2003adatom}& -1.37 & --- & 2.24 & -2.34 & --- & 2.20 \\
 Empirical potential \eqref{eq1}& -1.86 & 5.20 & 2.31 & -1.96 & 4.88 & 2.25 \\
\end{tabular}
\end{ruledtabular}
\end{table*}

The results of our \textit{ab initio} calculations are in reasonable agreement with the data from Ref~\cite{stengel2003adatom}. Some discrepancy in the values of the binding energies $W_{0}$ and $W_{0}^{vac}$ is probably due to the differences in the calculation models used.

Estimating the potential parameters by formulas \eqref{eq13} and \eqref{eq14} for a substrate without a vacancy gives the values $\varepsilon = 2.67 \times 10^{-2}$~eV and $\sigma = 2.70$~\AA. Substituting these parameters into expression \eqref{eq11}, we obtain for a substrate with a vacancy: $W_{0}^{vac} = -1.46$~eV and $H_{0}^{vac} = 4.77$~\AA. Hence, the proposed analytical model only partially reproduces the effects associated with changes in the energy $W_{0}$ and distance $H_{0}$ due to the formation of a vacancy on the substrate surface. The analytical model slightly underestimates the distance $H_{0}$ compared to the DFT, probably due to the simplified vacancy form. The binding energy in absolute value is strongly underestimated (1.46~eV vs 2.04~eV).

The calculation using the LAMMPS program with the specified empirical force field does not predict a valuable change of $W_{0}$ and $H$ due to the vacancy. Both with or without a vacancy, the optimization gives the values of $W_{0} \approx -0.9$~eV and $H_{0} \approx 5.4$~\AA. Thus, the model gives the correct values for the ideal $\mathrm{Al}(111)$ surface but does not predict the changes associated with the appearance of a vacancy.

Based on expressions \eqref{eq11} and \eqref{eq12}, one can assume that a change in the parameter $\varepsilon$ with a constant $\sigma$ should not affect the value of $H_{0}$, and $W_{0}$ is linearly dependent on  $\varepsilon$. However, the optimization using the LAMMPS program indicates the opposite. The origin of the discrepancy is that when deriving equations \eqref{eq11} and \eqref{eq12}, we assumed the ideal rigidity of the substrate and sphere. In numerical calculations, we used realistic potentials for the carbon and aluminum atoms \cite{tersoff1989modeling, sheng2011highly}, allowing for the displacement of the atoms and the deformation of the surfaces. Substituting $W_{0}^{vac} =-2.04$~eV and $H = 5.50$~\AA\ from this work into expressions \eqref{eq13} and \eqref{eq14}, we obtain $\varepsilon = 5.27 \times 10^{-2}$~eV and $\sigma = 2.70$~\AA. The optimization in the LAMMPS program using the above parameters gives $H_{0} = 5.3$~\AA\ for the case without a vacancy and $H_{0}^{vac} = 4.9$~\AA\ for the case with a vacancy. This change in the position of the fullerene is in reasonable agreement with the results of DFT calculations. The value of $W_0$, when calculated using the potential \eqref{eq1}, change much less than the value estimated \textit{ab initio}. Without a vacancy, the binding energy calculated in the LAMMPS program was $W_{0} = -1.86$~eV, and with a vacancy $W_{0}^{vac} =-1.96$~eV. Hence, when using the potential \eqref{eq1} with the proposed parameters to calculate the interaction of $\mathrm{C}_{60}$ with an ideal $\mathrm{Al}(111)$ substrate the binding energy turns out to be highly overvalued. We assume that this is not essential for the study of $\mathrm{Al}-\mathrm{C}_{60}$ composite materials since the existing technologies for the production of bulk composites involve high-energy processing of the substance, which activates the processes associated with the reconstruction of the interphase.

Substituting $W_{0}^{vac} =-2.04$~eV and $H_{0}^{vac} = 4.99$~\AA\ into formulas \eqref{eq13} and \eqref{eq14} gives $\varepsilon = 0.175$~eV and $\sigma = 1.99$~\AA. The use of these parameters when optimizing the atomic positions in the LAMMPS program leads to a significant distortion of the substrate structure, which results in doubtful values of $W_{0}$ and $H_{0}$ obtained from the empirical simulations.

The analysis allows us to conclude that the best correspondence with the results of DFT calculations gives the potential \eqref{eq1} with the parameters $\varepsilon = 5.27 \times 10^{-2}$~eV and $\sigma = 2.70$~\AA. Therefore, below we used these values. Comparing the theoretical estimates with the experimental data \cite{maxwell1998electronic, hamza1994reaction} also indicates the preferred use of these potential parameters (see Section~\ref{sec:tpd}). We also estimated the data shown in Table~\ref{tab:tab1} using these values. Table~\ref{tab:tab1} shows that the data obtained by using the potential \eqref{eq1} for the aluminum surface with a vacancy are in better agreement with \textit{ab initio} calculations than the results for an ideal slab. Note that the potential \eqref{eq1} with the applied parameters significantly overestimates the binding energy for fullerenes and the ideal $\mathrm{Al}(111)$ slab. The primary origin of the differences in the binding energies of $W_{0}$ and $W_{0}^{vac}$ is the dependence of the chemical activity of aluminum atoms on their local environment. The potential \eqref{eq1} does not consider this effect. The change in the position of the fullerene relative to the substrate is due to the finite atomic volume and is accurately reproduced using the potential \eqref{eq1}.

Note that the use of the pair potential (1) makes it impossible to take into account the redistribution of a charge during the formation of a polar covalent $\mathrm{Al} - \mathrm{C}$ bond. Also, the change in the interaction parameters for pairs of $\mathrm{Al} - \mathrm{Al}$ and $\mathrm{C} - \mathrm{C}$ atoms, caused by the redistribution of the electron density during the formation of the $\mathrm{Al} - \mathrm{C}$ bond and being the reason for a change in the local structure of fullerene \cite{stengel2003adatom}, is not taken into account. The use of existing models, which approximately take into account the charge transfer during the formation of a chemical bond and the dependence of its energy on the local environment of atoms (for example \cite{liang2013classical,liang2013reactive}), is limited due to the lack of parameterization with the $\mathrm{Al} - \mathrm{C}_{60}$ systems included in the training set.

\subsection{\label{sec:tpd}Kinetics of fullerene desorption from an aluminum surface at a linear change of temperature over time.}

Consider the process of desorption of fullerenes from the surface of a substance when the system is heated from temperature $T_{0}$ to $T_{1}$ at a constant rate $c$,

\begin{equation}\label{eq16}
    T = T_{0}(1 + c t).
\end{equation}

To develop an analytical model, we used the following simplifying approximations. We considered the fullerenes as point particles, neglecting the internal degrees of freedom and rotational energy. The in-plane motion of the particles was assumed free and unlimited. This assumption follows from the experiments \cite{maxwell1998electronic, johansson1998adsorption}, where the high mobility of fullerenes in the substrate plane was observed even at room temperature. The desorption kinetics was investigated using the transition state theory. The activation free energy was calculated with the harmonic approximation and assumed to be equal to the difference between the free energies of the reaction products and the reactants. Ignoring adsorption, we supposed that the fullerenes removed from the surface could no longer interact with the substrate.

The applied simplifications allow us to estimate the desorption rate using the equation \cite{king1975thermal}:

\begin{equation}\label{eq17}
    k = -\frac{n \omega}{2 \pi} \exp\left(-\frac{\Delta F}{k_{B} T} \right),
\end{equation}

\noindent where $k = d n/ d t$ is the desorption rate, $n$ is the coverage density (the number of molecules per unit surface), $\Delta F$ is the activation free energy, $\omega$ is the vibrational frequency of the molecule in the direction of the normal to the surface and $k_{B}$ is the Boltzmann constant.

Expression \eqref{eq17} uses the Boltzmann distribution, which imposes certain restrictions on the conditions of the problem. The desorption relaxation time must be much longer than the temperature relaxation time of the system perturbed by the elementary reaction act. In our model, we assume that the requirement is fulfilled.

The solution of equation \eqref{eq17} has the following general form:

\begin{equation}\label{eq18}
    n(t) = n_{0} \exp \left[ -\int_{0}^{t} \frac{\omega}{2 \pi} \exp \left(-\frac{\Delta F}{k_{B} T} \right) d t \right].
\end{equation}

Using the harmonic approximation, one can calculate the configuration integral analytically and obtain the expression for the activation free energy

\begin{equation}\label{eq19}
    \Delta F = -W_{0} + \frac{1}{2} k_{B} T \ln \left( \frac{2 \pi k_{B} T}{\alpha \sigma^{2}} \right).
\end{equation}

\noindent Here the stiffness $\alpha$ is determined by the value of the second derivative of the potential energy from \eqref{eq12} and is related to the frequency and the mass of the oscillator as follows:

\begin{equation}\label{eq20}
    \omega = \sqrt{\frac{\alpha}{m}} = \left(\frac{1}{m} \frac{d^{2} W}{d H^{2}} \right)^{1/2}.
\end{equation}

Substituting $W(H)$ from \eqref{eq12} into \eqref{eq20}, we obtain

\begin{equation}\label{eq21}
    \alpha = 4 \pi^{2} n_{1} n_{2} R \sigma^{2} \varepsilon \left[\frac{2}{5} (\zeta_{10}^{-} - \zeta_{10}^{+}) - (\zeta_{4}^{-} - \zeta_{4}^{+}) \right].
\end{equation}

Only the temperature depends on time in the integrand of the right-hand side of \eqref{eq18} in the accepted formulation. The authors of Refs~\cite{maxwell1998electronic, johansson1998adsorption} performed annealing at a constant temperature. In this case, the integral under the sign of the exponent gives a linear time function, and the following formula determines the evolution of the coverage density:

\begin{equation}\label{eq22}
    n(t) = n_{0} e^{-t/\tau},
\end{equation}

\noindent where the relaxation time $\tau$ is the constant value, which has the form:

\begin{equation}\label{eq23}
    \tau = 2 \pi \omega^{-1} \left( \frac{\alpha \sigma^{2}}{2 \pi k_{B} T} \right)^{-1/2} \exp \left( -\frac{W_0}{k_{B} T} \right).
\end{equation}

Let us consider the case of linear heating of the system, which Hamza et al. implemented in their studies \cite{hamza1994reaction}. After integration, we obtain the expressions for the coating density and the desorption rate

\begin{eqnarray}
    \label{eq24}
    n \approx && n_{0} \exp \left \{ -\frac{\omega}{ 2\pi c} \sqrt{-\frac{W_{0} \alpha \sigma^{2}}{2 \pi k_{B}^{2} T_{0}^{2}}} \left[ \psi (T) - \psi (T_{0})\right] \right \},
    \\
    \label{eq25}
    k = && \frac{n \omega}{2 \pi} \sqrt{\frac{\alpha \sigma^{2}}{2 \pi k_{B} T}} \exp \left( \frac{W_{0}}{k_{B} T} \right).
\end{eqnarray}

\noindent In expression \eqref{eq24} we used only the first two terms of the asymptotic expansion of the additional error function $\erfc(x)$ for large arguments $x = |W_{0}| / (k_{B} T) \gg 1$ and applied the notation:

\begin{equation}\nonumber
    \psi(T) = \left( -\frac{W_{0}}{k_{B} T} \right)^{-3/2} \exp \left(\frac{W_{0}}{k_{B} T} \right)
\end{equation}

It can be shown that for a small heating rate, equation \eqref{eq24} takes the form

\begin{equation}\label{eq26_lim}\nonumber
    n \approx n_{0} \exp \left[\frac{\omega t}{2 \pi} \sqrt{\frac{\alpha \sigma^{2}}{2 \pi k_{B} T_{0}}} \exp \left( \frac{W_{0}}{k_{B} T_{0}} \right) \left( 1 - \frac{3 k_{B} T_{0}}{2 W_{0}} \right) \right],
\end{equation}

\noindent which coincides with formulas \eqref{eq22} and \eqref{eq23} up to a small additive to the relaxation time $\delta \tau \propto 3 k_{B}T_{0} / (2 W_{0})$. 

\begin{figure}
\includegraphics[width=1\linewidth]{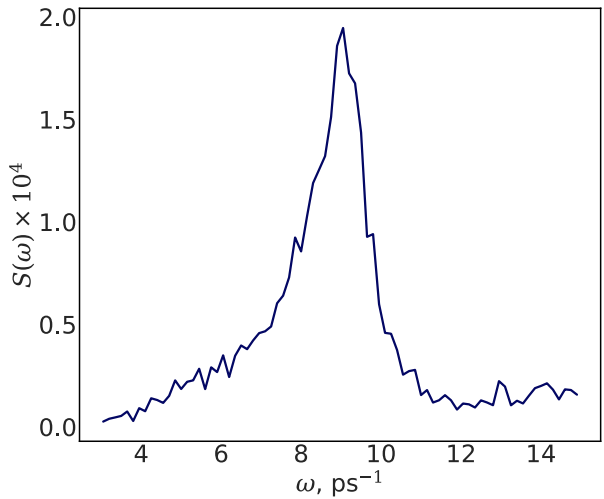}%

 \caption{\label{fig:fig3}%
  Vibrational spectral density of the centers of mass of fullerenes in the normal direction relative to the aluminum substrate.
 }%
\end{figure}

Using the data from Table~\ref{tab:tab1} and the corresponding values of the parameters $\varepsilon$ and $\sigma$ we calculated the stiffness $\alpha$ and frequency $\omega$ according to equations \eqref{eq20} and \eqref{eq21}: $\alpha = 9.11$~eV\AA$^{-2}$ and $\omega = 10.8$~ps$^{-1}$. To verify the analytical estimates and analyze the model's applicability, we performed a spectral analysis of the trajectories of the centers of mass of fullerenes. The calculated spectral density of the oscillations is shown in Fig.~\ref{fig:fig3}.

The maximum of the function $S(\omega)$ corresponds to the frequency $\omega = 9.1$~ps$^{-1}$, which is within 20\% coincides with the analytical value. Considering the significance of the accepted assumptions, we regard such coincidence of the results as acceptable. Note that using a slightly overvalued oscillation frequency in the analytical calculation may overestimate the desorption rate.

The estimation of the desorption relaxation time by formula \eqref{eq23} at a temperature of 730~K gives $\tau \approx 5.3$~s. Using the binding energy $W_{0} = -2.34$~eV from Ref~\cite{stengel2003adatom}, we obtain $\tau \approx 544$~s. Considering the experimental conditions in \cite{maxwell1998electronic, johansson1998adsorption}, one can conclude that the obtained relaxation times seem quite realistic. Using $W_{0} = -0.98$~eV, estimated for the ideal $\mathrm{Al}(111)$ substrate, one obtains too short relaxation time, $\tau \approx 5.3 \times 10^{-7}$~s. The analysis shows that at such binding energy, desorption of fullerenes from the aluminum surface would be observed due to annealing at a temperature of 350~K with a relaxation time of 8~s, which contradicts the experimental data from Refs~
\cite{maxwell1998electronic, johansson1998adsorption}.

The MD calculation limits us to a process of time about 1~ns, while the desorption time with the parameters defined above is much longer. Formally, by lowering the binding energy value $W_{0}$ and increasing the process temperature, we can significantly accelerate the desorption. Although the computational conditions will not correspond to the experimental ones, the comparison of the analytical results with the data from direct MD simulations is essential for testing both the analytical model and the applicability of the potential \eqref{eq1}.

We performed the MD simulations of the desorption for different temperatures and parameters of the interaction potential. In the first case, we elevated the temperature from 1000~K to 2200~K and set the value of $W_{0} = -1.43$~eV (70\% of the value calculated using DFT). This temperature range corresponds to the equilibrium of aluminum in the liquid phase. In the second case, we considered the crystal $\mathrm{Al}(111)$ substrate and performed the heating from 400~K to 800~K. In this instance, we reduced the binding energy by four times compared to the calculated one, $W_{0} = -0.51$~eV. In both cases, we considered the heating time of 1~ns. According to \eqref{eq13}, the value of the parameter $\sigma$ does not depend on $W_{0}$; therefore, we assumed that $\sigma = 2.70$~\AA.

\begin{figure*}
\begin{minipage}{0.49 \linewidth}
\includegraphics[width=1\linewidth]{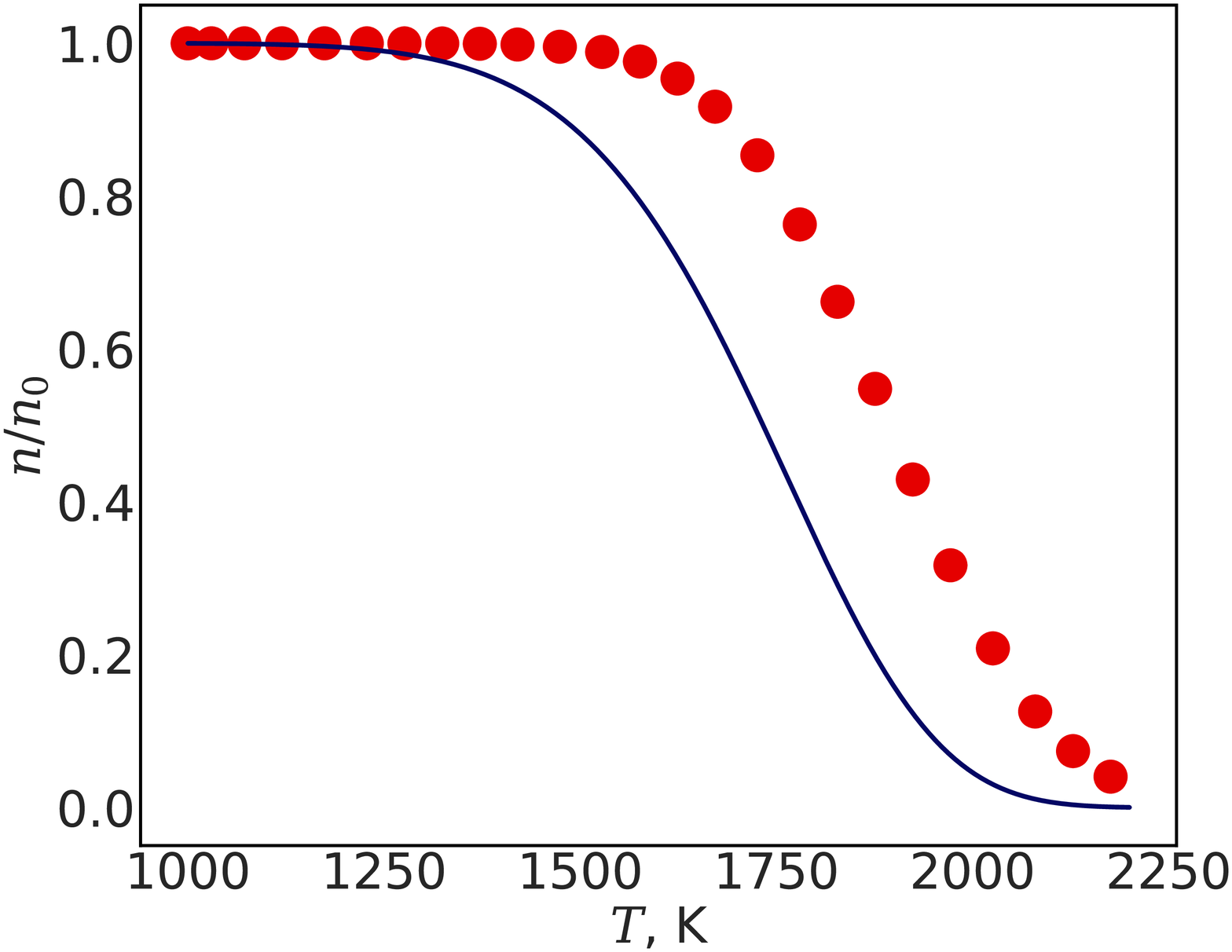} a%
\end{minipage}
\begin{minipage}{0.49 \linewidth}
\includegraphics[width=1\linewidth]{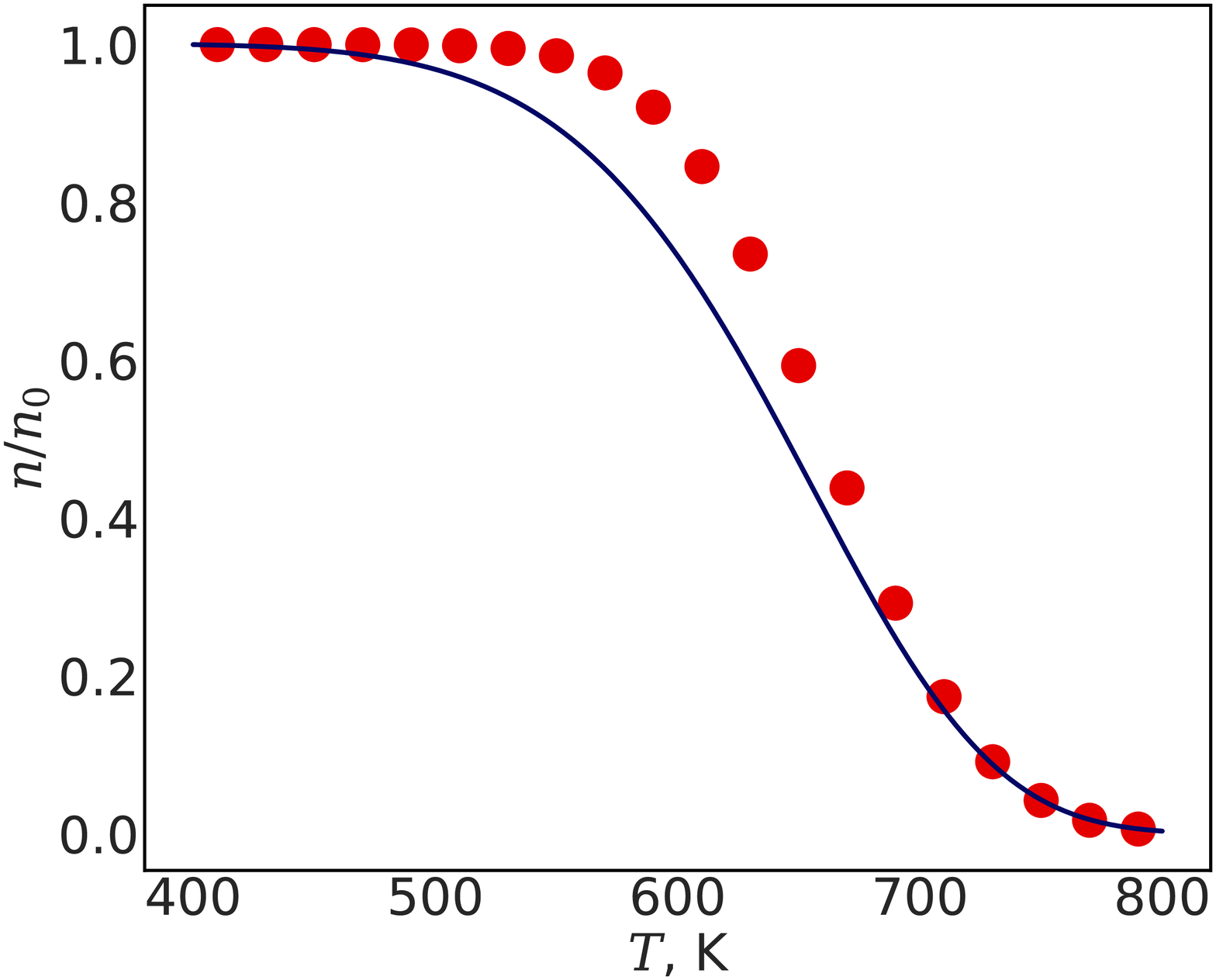} b%
\end{minipage}

 \caption{\label{fig:fig4}%
  Temperature dependences of the density of the aluminum substrates coating with fullerenes.  The solid lines correspond to the analytical estimations according with formula \eqref{eq24} and circles indicate the results from the MD simulations:
  (a)~-- the simulation was performed for the liquid aluminum film and $W_{0} = -1.41$~eV, and (b)~-- desorption from the solid slab $\mathrm{Al}(111)$ with $W_{0} = -0.51$~eV.
 }%
\end{figure*}

The temperature dependences of the coating densities calculated analytically and by the MD method under the specified conditions are shown in Fig.~\ref{fig:fig4}. Good agreement between the results of numerical and analytical estimations indicates the adequacy of the model. The analytical calculation slightly underestimates the desorption temperature in comparison with the numerical one. Partly, this effect is due to the previously discussed inexactness in analytical calculations of the vibration frequencies. Besides, we note that the disagreement is more pronounced for the melt than for the crystal substrate. An increase in the interaction between fullerenes and the liquid surface is associated with capillary phenomena.

Molecular dynamics simulation does not predict the formation of $\mathrm{Al}_{4}\mathrm{C}_{3}$ carbide at high temperatures, which was observed in experiments with fullerenes and carbon nanotubes \cite{evdokimov2021,aborkin_2021}. This is explained by the short times of molecular dynamics simulation and the activation nature of the chemical reaction.

\begin{figure*}
\begin{minipage}{0.49 \linewidth}
\includegraphics[width=1\linewidth]{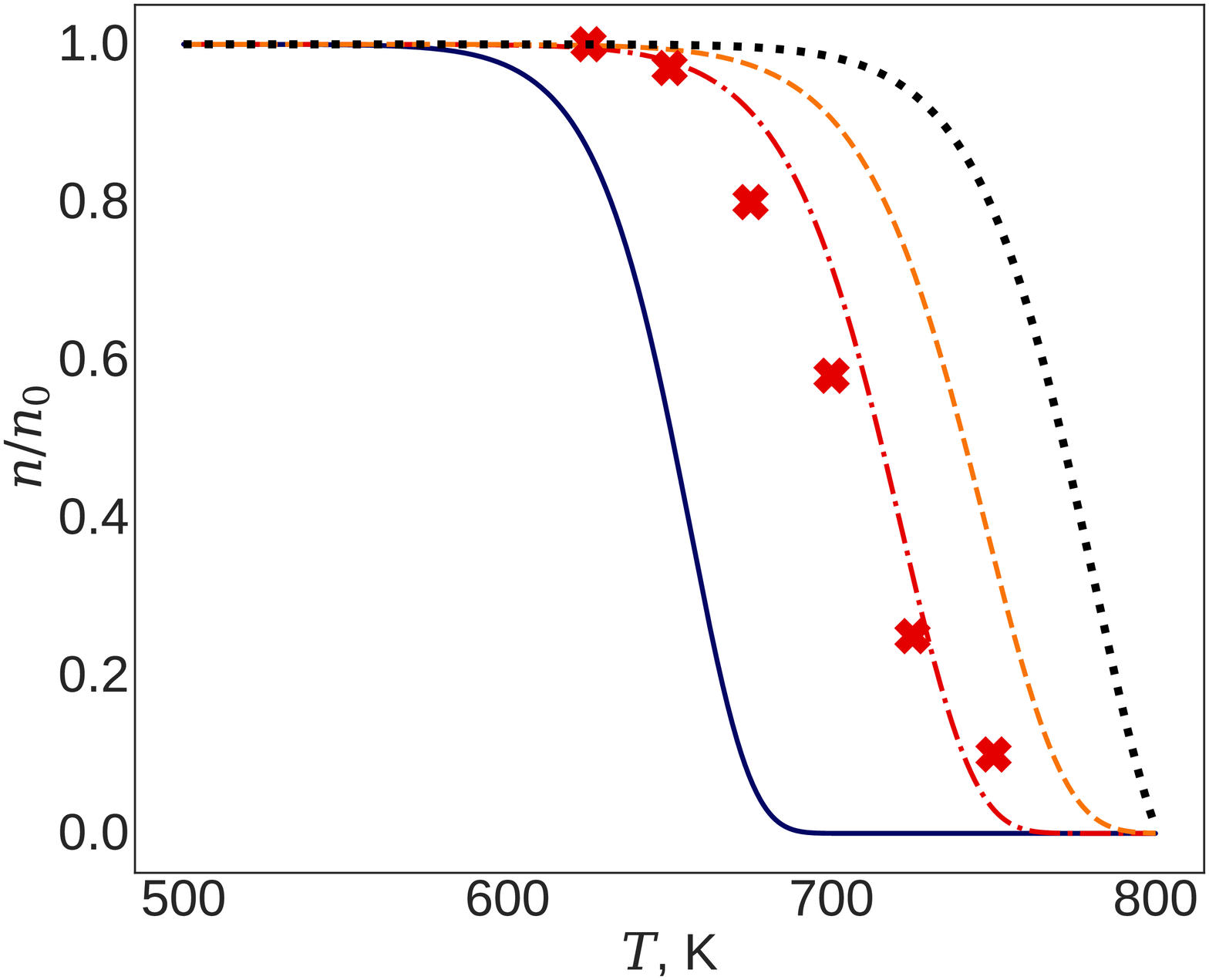} a%
\end{minipage}
\begin{minipage}{0.49 \linewidth}
\includegraphics[width=1\linewidth]{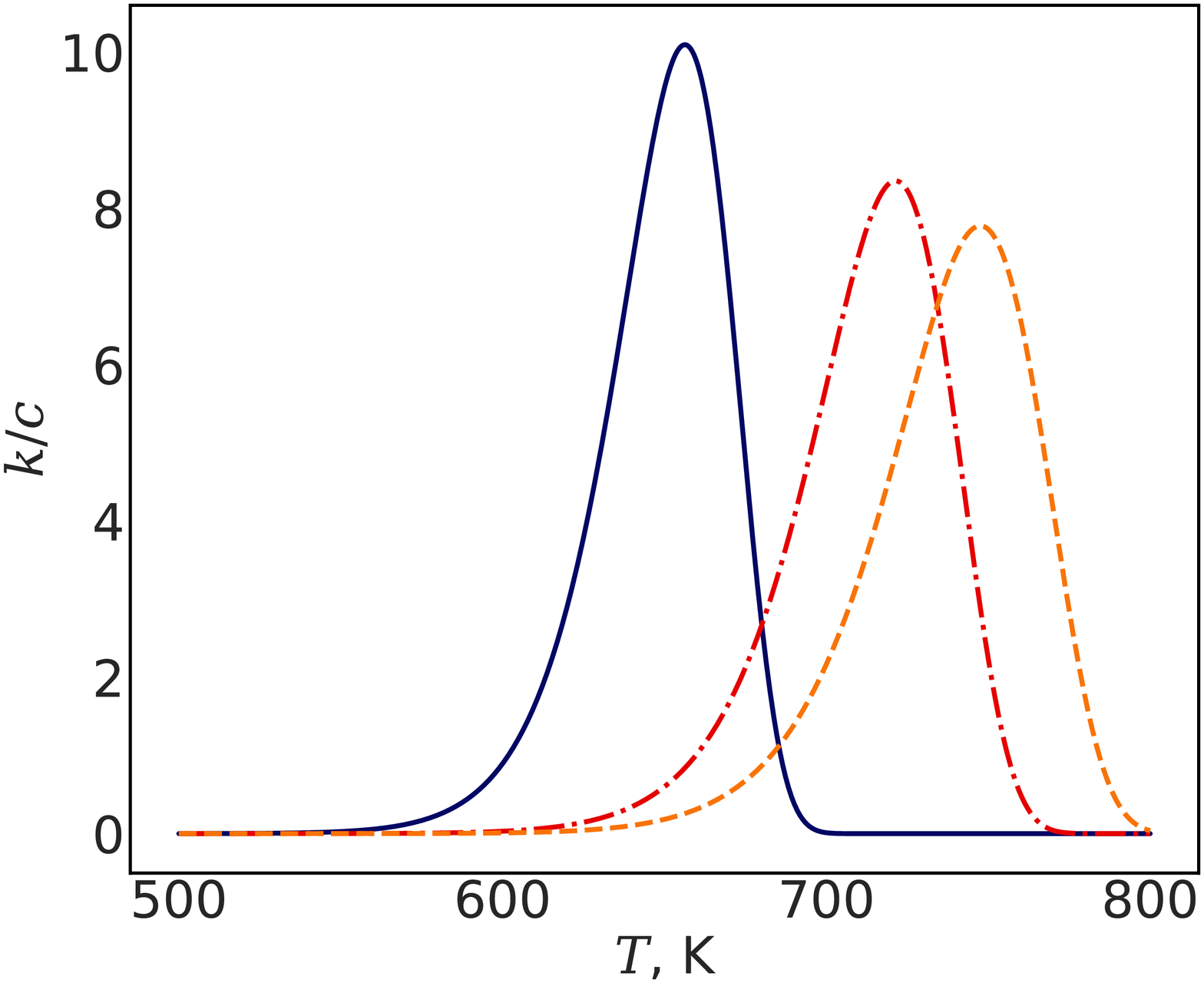} b%
\end{minipage}

 \caption{\label{fig:fig5}%
Temperature dependences of the surface coating densities (a) and the desorption rates (b) calculated by equations \eqref{eq24} and \eqref{eq25}. The solid line corresponds to the heating rate of 0.1~K/s; the dash-dotted line, to 3~K/s; and the dashed line, to 10 K/s. The experimental data \cite{hamza1994reaction} obtained at a heating rate of 3~K/s are labelled with the crosses. The dotted line shows the data from the analytical calculations obtained with the value of $W_{0} = -2.34$~eV from Ref~\cite{stengel2003adatom} }%
\end{figure*}

Under the conditions of TPD studies, the heating rates can vary widely \cite{king1975thermal}. We considered the heating rates $c \times T_{0} = 0.1$, 3 and 10~K/s [the variable designations correspond to \eqref{eq16}] to analyze the desorption kinetics. The theoretical dependences $n(T)$ were compared with the experimental data by Hamza et al. \cite{hamza1994reaction}, where they used a heating rate of 3~K/s. The results are shown in Fig.~\ref{fig:fig5}a. Moreover, we used the model to calculate the dependence $n(T)$ with the value of $W_{0} = -2.34$~eV, taken from Ref~\cite{stengel2003adatom}, and the heating rate of 3~K/s, corresponding to the experiment \cite{hamza1994reaction} (see Fig.~\ref{fig:fig5}a). The calculated temperature dependences of the desorption rate related to the different heating rates are shown in Fig.~\ref{fig:fig5}b.

Figure~\ref{fig:fig5} shows that the proposed analytical model reproduces the experimental data \cite{hamza1994reaction} with high accuracy when using the results from our \textit{ab initio} calculations. The use of the binding energy value estimated by Stengel et al. \cite{stengel2003adatom} slows down the desorption and worsens the coincidence between the theory and experiment. With an increase in the heating rate, the maximum of the desorption rate shifts to the region of higher temperatures. The maximum desorption rates at heating rates of 0.1, 3 and 10~K/s correspond to 685, 757 and 785~K, respectively. The obtained temperature range from 685~K to 785~K is entirely consistent with the data from Refs~\cite{maxwell1998electronic, johansson1998adsorption}, in which the authors reported the desorption temperature of 730~K. The simulation results indicate a single peak of the function $k (T)$, which is a consequence of the applied assumption about the constancy of the binding energy $W_{0}$, regardless of fullerenes' location on the substrate. 

\subsection{\label{sec:capillary}Capillary phenomena at the interface between molten aluminum and $\mathrm{C}_{60}$ fullerenes immersed in it}

Analysis of the state of the aluminum melt with fullerenes immersed in it indicates the system's disequilibrium. However, we can distinguish several processes with different characteristic times. The relaxation of the atomic velocity distribution function is the fastest process, the characteristic time of which is $\tau_{at} \sim 1-10$~ps, i.e. $10-100$ times greater than the period of atomic vibrations. The equilibration of the translational motion of fullerenes takes a little longer, $\tau_{\mathrm{C}_{60}} \sim 5 \tau_{at}$. One can roughly evaluate the time of this process accounting that in an head-on elastic collision of particles with masses $m_{1}$ and $m_{2}$ ($m_{1} < m_{2}$), the redistributed energy is $\Delta E \propto \sqrt{m_{1}/m_{2}}$. The relaxation of the spatial distribution of fullerenes is determined by the diffusion coefficient $D$ of nanoparticles in the melt. The value $D \sim 10^{-5}$~cm$^{2}$/s can be considered as a reasonable estimate. Thus, the relaxation time of the diffusion process is $\tau_{D} = 1/(4 D n_{s_{0}}^{2/3}) \sim 10^{-9}$~s, where $n_{s_{0}} \approx 1.15 \times 10^{-4}$~\AA$^{-3}$ is the concentration of fullerenes per unit volume. 

Besides, it is known from the theory of colloids that with the poor wettability by an environment, particles tend to coagulation due to capillary forces \cite{smoluchowski1918versuch, derjaguin1987surface, botto2012capillary}. The process of clusterization of fullerenes in an aluminum melt was also observed in our calculations. For a preliminary evaluation of the process's characteristic time, we used the assumptions from the Smoluchowski theory \cite{smoluchowski1918versuch, derjaguin1987surface}. Let us assume that the ensemble of fullerenes can be represented as an ideal gas of Brownian particles. We also assumed that the elementary act of coagulation always occurs when two fullerenes get closer to each other at a distance less than or equal to $2 R_{f} \approx 10.4$~\AA. To describe the process, we used the kinetic equation

\begin{equation}\label{eq27}
    \frac{d n_{s}}{d t} = -k_{C} n_{s}^2,
\end{equation}

\noindent where $n_{s}$ is the concentration of free fullerenes, which was calculated directly during the MD simulation, initially $n_{s} = n_{s_{0}}$. The coagulation rate constant $k_{C}$ in Smoluchowski theory is defined as

\begin{equation}\label{eq28}
    k_{C} = 8 \pi R_{f} D.
\end{equation}

The linear time dependence of the inverse concentration of free fullerenes follows from equation \eqref{eq27} 

\begin{equation}\label{eq29}
    \frac{n_{s_{0}}}{n_{s}} = 1 + n_{s_{0}} k_{C} t.
\end{equation}

The time $\tau_{1/2}$, corresponding to a halving of the free particle concentration, is commonly used as the characteristic coagulation time. Using equations \eqref{eq28} and \eqref{eq29} one can assesse the coagulation time as $\tau_{1/2} = (8 \pi n_{s_{0}} R_{f} D)^{-1} \sim 7 \times 10^{-10}$~s.

The evaluated time scales of the relaxation processes suggest a reasonable choice of the MD simulation time of 10~ns. According to the above assessment based on the assumption of absent interaction between the fullerenes in the melt, the diffusion and coagulation times are of the same order of magnitude. Therefore, it is impossible to separate the processes into ``fast'' and ``slow'' ones. However, the diffusion and coagulation rates from MD simulations differ significantly, which indicates a strong interaction between the fullerenes in the melt and allows us to separate the processes by the characteristic times.

\begin{figure}
\includegraphics[width=1\linewidth]{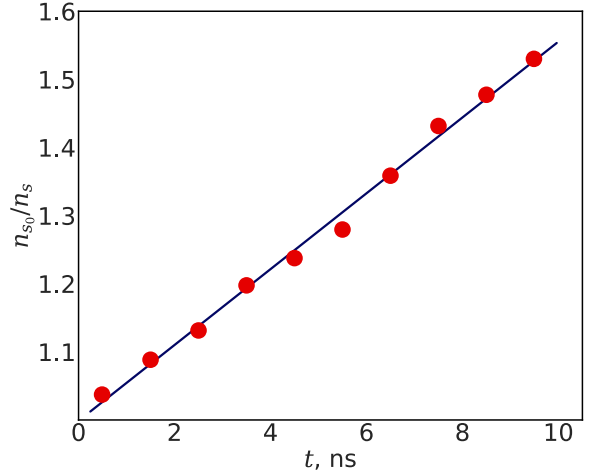}%

 \caption{\label{fig:fig6}%
  Time dependence of the inverse concentration of free fullerenes: the circles indicate the results of the MD calculations, and the solid line is an approximation with \eqref{eq29}, where the $k_{C} \approx 4.84 \times 10^{-13}$~cm$^{3}$~s$^{-1}$ is the coagulation rate constant.
 }%
\end{figure}

In this paper, we calculated the coagulation rate more accurately, analysing the trajectories of the centers of mass of fullerenes from the MD simulations. The coagulation rate constant was estimated by fitting the time dependence of the concentration of free fullerenes with formula \eqref{eq29}. We assumed fullerenes bound if the distance between their centers of mass did not exceed $2 (R + R_{c}) \approx 9.5$~\AA, where $R_{c} = 2.1$~\AA\ is the cutoff radius of the Tersoff potential \cite{tersoff1989modeling,tersoff1988}. The time dependence of the inverse concentration of free molecules is shown in Fig.~\ref{fig:fig6}.

\begin{figure*}
\begin{minipage}{0.49 \linewidth}
\includegraphics[width=1\linewidth]{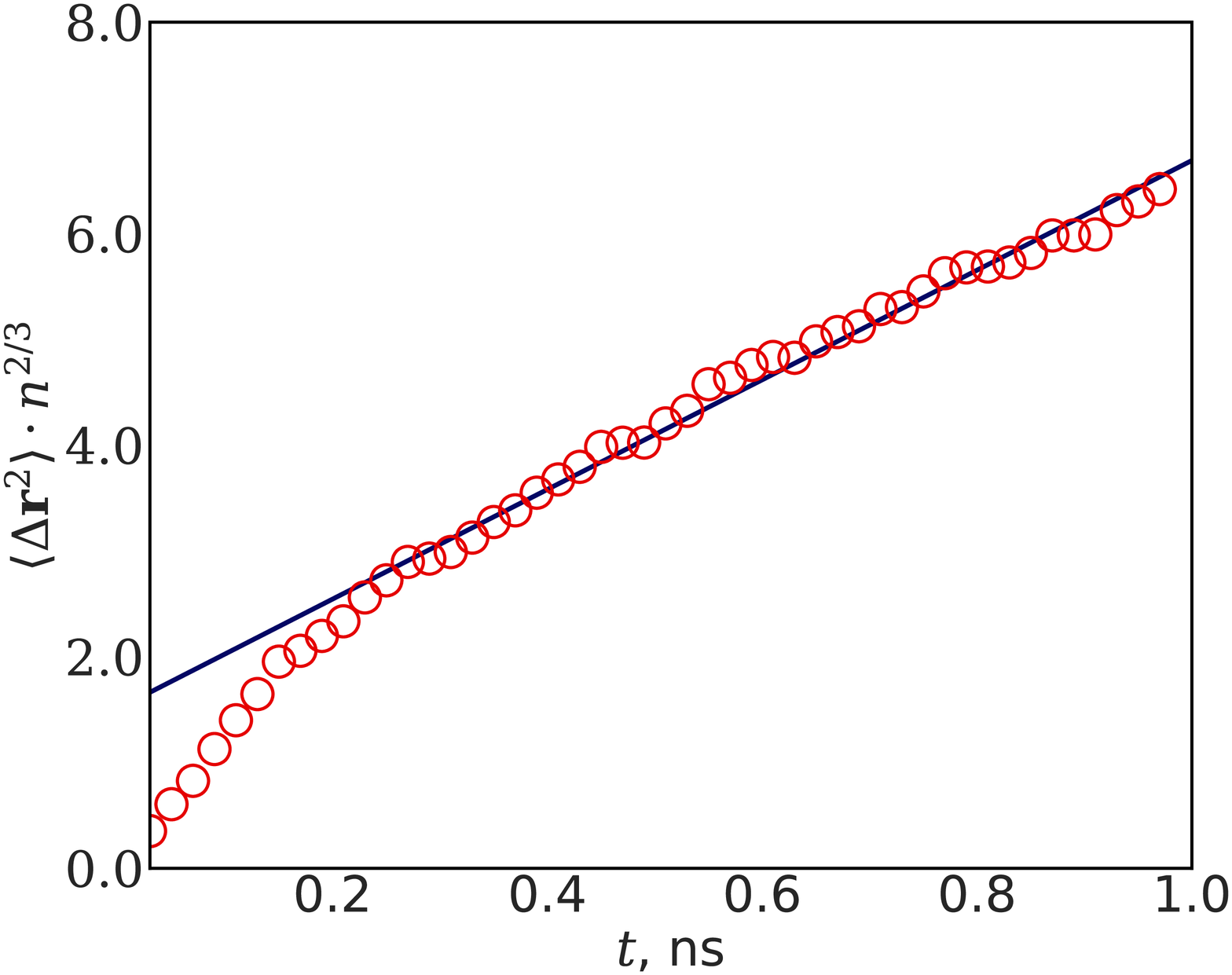} a%
\end{minipage}
\begin{minipage}{0.49 \linewidth}
\includegraphics[width=1\linewidth]{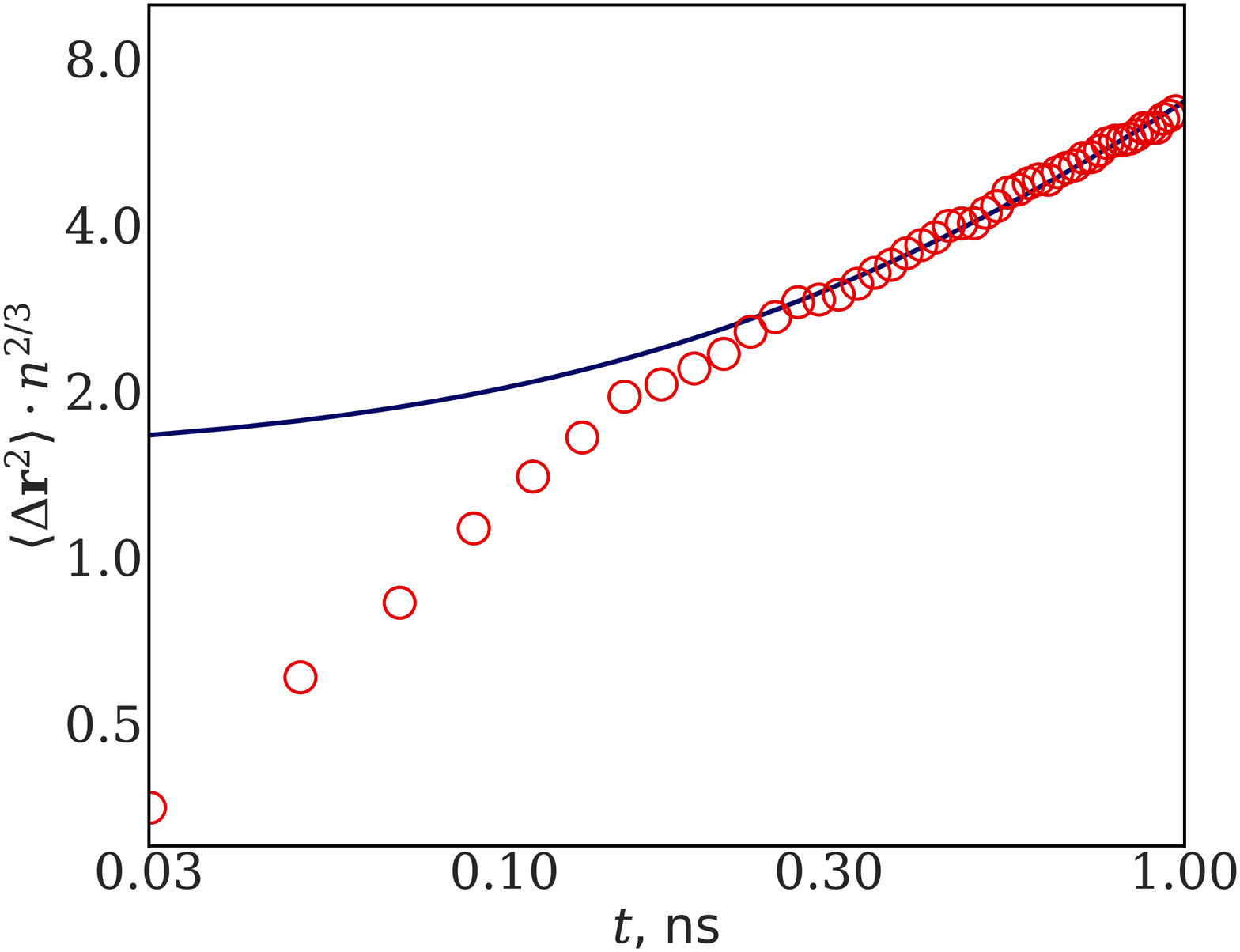} b%
\end{minipage}

 \caption{\label{fig:fig7}%
Time dependence of MSD plotted in linear (a) and logarithmic (b) scales. Circles indicate results of MD calculations, and the solid line is the linear approximation. }%
\end{figure*}

The characteristic time of the coagulation process $\tau_{1/2} = (k_{C} n_{s_{0}})^{-1} \approx 18.0$~ns is about 25 times higher than the value assessed with the Smoluchowski theory. On the time intervals $t \ll \tau_{1/2}$ the coagulation can be neglected. Only about 5\% of fullerenes formed connected clusters during the first nanosecond of the simulations. Therefore, for the time interval $t \leq 1$~ns, the diffusion coefficient of free fullerenes in the melt can be calculated. Figure~\ref{fig:fig7} shows the time dependence of the mean square displacement (MSD) of fullerenes at $t \leq 1$~ns.

There are two characteristic modes of the time dependence of MSD in simple liquids. For the ballistic mode, the dependence has a form of $\langle \Delta r^2 \rangle = v^2 t^2$ and for the diffusion regime, $\langle \Delta r^2 \rangle = 6 D t$. For separation of these modes, it is convenient to use the plot on a logarithmic scale. Figure~\ref{fig:fig7} shows that the diffusion regime corresponds to a time $t > 0.2$~ns. We performed the linear approximation for this time interval. The diffusion coefficient $D = 3.64 \times 10^{-5}$~cm$^2$/s, obtained from the MD simulations, coincides by order of magnitude with the value $D \sim 10^{-5}$~cm$^2$/s, which we used above in our evaluations. The obtained results allow us to refine the diffusion relaxation time, $t_{D} = 0.29$~ns. Recalculation of the coagulation rate constant by the Smoluchowski formula \eqref{eq28} gives $k_{C} \approx 4.77 \times 10^{-11}$ ~ cm$^{3}$~s$^{-1}$, which corresponds to the characteristic time $\tau_{1/2} \approx 0.18$~ns.

The comparison of relaxation times confirms the possibility of dividing the nonequilibrium processes of coagulation and diffusion into ``slow" and ``fast" ones. We emphasize that the assessment of the coagulation rate constant from the Smoluchowski theory, assuming no interactions between the fullerenes, gives the value approximately 100 times higher than the obtained from the MD simulations. Therefore, we conclude that the interaction between fullerenes is nonzero at distances greater than $2 (R + R_{c})$ and the coagulation process is of an activation nature.

\begin{figure}
\includegraphics[width=1\linewidth]{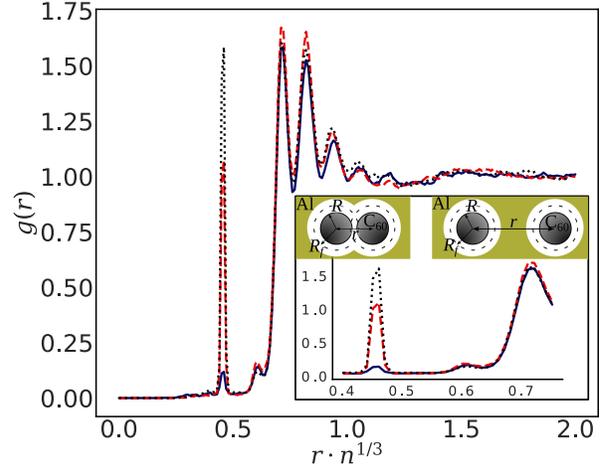}%

 \caption{\label{fig:fig_8}%
    RDFs of the centers of mass of the fullerenes immersed in the aluminum melt. The solid line indicates the RDF averaged on the time interval $t \in [0, 1]$~ns; the dashed line, $t \in [5, 6]$~ns; and the dotted line, $t \in [9, 10]$~ns. The insert illustrates the relative positions of pairs of fullerenes for the first two peaks of the RDFs.}%
\end{figure}

Figure~\ref{fig:fig_8} shows the radial distribution functions (RDFs) of the centers of mass of the $\mathrm{C}_{60}$ molecules. The functions were averaged over different time intervals: at $t \in [0, 1]$~ns, $t \in [5, 6]$~ns, and $t \in [9, 10]$~ns.

The evolution of the RDF is associated with the growth of a narrow peak in the vicinity of the point of $r \cdot n^{1/3} = 0.45$ (9.24~\AA), which characterizes the number of bound molecules. Function $g(r)$ has pronounced maxima at points 0.71, 0.82 and 0.94 (14.6, 16.8, and 19.3~\AA), which indicates the nonideality of the ensemble of fullerenes. In our simulations, we apply only the short-range potential for the interaction between carbon atoms. The cutoff radius of the Tersoff potential \cite{tersoff1989modeling,tersoff1988} is 2.1~\AA. If the distance between the centers of mass is higher than 9.5~\AA, the Tersoff potential does not contribute to intermolecular interaction. The interaction between fullerenes at large distances is presumably due to the elastic deformation of the aluminum melt by capillary forces.

We emphasize that the present work does not take into account the charge redistribution during the formation of polar covalent $\mathrm{Al} - \mathrm{C}$ bonds, and fullerenes are considered electrically neutral. We assume that taking into account the Coulomb repulsion between fullerenes will slow down the coagulation process; therefore, the value of the rate constant calculated in this work can be considered to be an upper bound. Investigation of the effect of Coulomb corrections to the potential of intermolecular interaction on the rate of coagulation of fullerenes in a melt is beyond the scope of this work and can be considered elsewhere.

The change in the state parameters resulting from the system's relaxation during the formation of bound clusters of fullerenes is insignificant. The differences between the temperature and pressure values calculated and averaged for the first and tenth nanoseconds are 3~K and 300~bar, and the standard deviations are 0.85~K and 80~bar. Thus, the state parameters are approximately constant during the relaxation.

The calculation of the Laplace pressure in these conditions gives a value of $p = 1.49$~GPa. This value changes slowly during the coagulation of fullerenes: a decrease in the simulation after 10~ns did not exceed 0.5\%. A positive value of $p$ indicates the poor wettability of fullerenes by the melt. The surface tension calculated with formula \eqref{eq6} is $\gamma = 0.39$~J/m$^2$. We estimated the surface tension of aluminum with the same EAM potential in our previous paper \cite{reshetniak2020aluminumcarbon}. At 1000~K, the result was 0.71~J/m$^2$, which is higher by 0.32~J/m$^2$ than the value of $\gamma$ from the current study. Besides, the specific free energy of the interface between the aluminum melt and a graphite substrate was calculated in Ref~\cite{reshetniak2020aluminumcarbon}. The resulting value was lower by about 0.21~J/m$^2$ than $\gamma$. As the attraction between the carbon and aluminum atoms for fullerenes is much stronger than for graphite (see ref~\cite{reshetniak2020aluminumcarbon}), it is somewhat surprising that the specific free energy of the interface is higher for $\mathrm{Al}-\mathrm{C}_{60}$. Seemingly, this effect can be related to the spherical geometry of the fullerene and the smaller number of atoms involved in the interaction.

\begin{figure}
\includegraphics[width=1\linewidth]{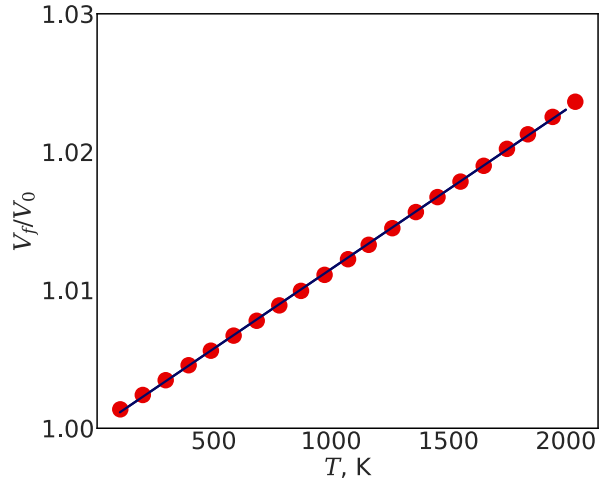}%

 \caption{\label{fig:fig9}%
Temperature dependence of the volume of a fullerene. Circles indicate the results from MD simulations, and the solid line is the linear approximation $V/V_{0} = \beta T + 1$; $V_{0} = 589.79$~\AA$^3$ is the volume of fullerene at 0~K and $\beta$ is the CTE.}%
\end{figure}

We considered an ensemble of 1000 noninteracting molecules to determine the CTE of fullerenes. The calculated temperature dependence of the volume of a fullerene is close to linear (see Fig.~\ref{fig:fig9}).

The CTE calculated by formula \eqref{eq7} using a linear approximation of the MD results was $\beta = 1.15 \times 10^{-5}$~K$^{-1}$. Our results differ from the data estimated by Kwon et al. \cite{kwon_thermal_2004}, where the authors obtained a nonlinear dependence $V(T)$ in the temperature range from 0 to 400~K, and the CTE took values from $-1 \times 10^{-5}$ to $2 \times 10^{-5}$~K$^{-1}$. Note that the authors obtained a strong dependence $\beta(T)$ at low temperatures. When the temperature elevated above 300~K, the CTE slowly increased, asymptotically approximating the constant value: $\beta \approx 2.0 \times 10^{-5}$~K$^{-1}$. The value of CTE estimated in this paper is noticeably lower than that calculated in \cite{kwon_thermal_2004}, which illustrates the previously discussed fact of the strong dependence of the parameters $\beta$, $\gamma$, $K$, etc., on the method for calculating the fullerene radius. Recall that we used the radius of the interface $R_{f} = R + 2^{-5/6} \sigma$ in this paper, and in \cite{kwon_thermal_2004} the authors used the average distance $R$ from the atoms to the center of mass of a fullerene. The recalculation of the CTE using $R$ gives a value of $\beta = 1.63 \times 10^{-5}$~K$^{-1}$ reasonably coinciding with the data from Ref~\cite{kwon_thermal_2004} at temperatures above 250~K.

Among the probable reasons for the discrepancy in the results at lower temperatures is the use of the Tersoff potential to describe the interaction between carbon atoms. Comparison with DFT shown that the Tersoff potential unsatisfactorily describes the thermal expansion of carbon nanoparticles at low temperatures \cite{kwon_thermal_2004}.

Due to the Laplace pressure of $1.49$~GPa, the fullerenes immersed in aluminum decrease their volume. Originally at 1000~K the fullerene radius $R_{f}$ was 5.223~\AA\ ($R = 3.705$~\AA), and after it was immersed in the melt, $R_{f} = 5.207$~\AA\ ($R = 3.689$~\AA). Calculation by formula \eqref{eq8} yields the bulk modulus $K_{T} = 162$~GPa, which differs from the results of DFT calculations \cite{khabibrakhmanov_2020, peon2014} (868~GPa and 874~GPa, respectively). In addition, there is a significant discrepancy between the results and the data of \cite{kaur2010}, where values in the range from 370 GPa to 694 GPa were obtained using different empirical potentials for carbon. The indicated discrepancies can only partly be explained by the method chosen for calculating the fullerene volume. The use of the mean distance from the center of mass to atomic nuclei as the fullerene radius in the calculations by formulas \eqref{eq5} and \eqref{eq8} gives the bulk modulus $K_{T} = 324$~GPa. This value is much lower than the value of 674~GPa obtained using the Tersoff potential in \cite{kaur2010}. The discrepancy is explained by the fact that in this work we used the original parametrization of the Tersoff potential proposed in \cite{tersoff1989modeling,tersoff1988}, while Kaur et al. \cite{kaur2010} modified the potential parameters in order to better describe the available experimental data. A somewhat better value of $K_{T} = 324$~GPa calculated in this work agrees with the value of 370~GPa from \cite{kaur2010}, obtained using the potential \cite{brenner1990}. Note that in our work the bulk modulus was calculated at 1000 K, while in \cite{kaur2010} the calculations were performed at zero temperature. Assuming a monotonic decrease in the values of elastic moduli with increasing temperature, we consider the indicated correspondence reasonable. The reasons for the discrepancy between the obtained results and the DFT simulation data can be associated with both an approximate description of interatomic interactions using the potential \cite{tersoff1989modeling,tersoff1988}, and with an approximate allowance for the electron correlation in DFT.

\section{\label{sec:conclusions}Conclusions}

The interaction between the surfaces of aluminum in liquid and solid states and $\mathrm{C}_{60}$ fullerenes was studied. We calculated the binding energy of fullerenes with the $\mathrm{Al}(111)$ substrate using DFT. The parameters of the Lenard-Jones potential for pairs of $\mathrm{Al} - \mathrm{C}$ atoms at the interface were determined. Analytical and numerical models of TPD of fullerenes from the $\mathrm{Al}$ surfaces were proposed. We took all empirical parameters used in the model from the \textit{ab initio} calculations. The analytical model yields the results that are in good agreement with the data of MD simulations and experiments from Refs~\cite{maxwell1998electronic, johansson1998adsorption, hamza1994reaction}.

We applied the potential \eqref{eq1} with the proposed parametrization to investigate the capillary phenomena at the interface between the aluminum melt and the fullerenes. The Laplace pressure affecting fullerenes at a temperature of 1000~K was $1.49$~GPa. Using the calculated Laplace pressure and the interface radius, we estimated the surface tension $\gamma = 0.39$~J/m$^{2}$. This value is higher than the surface tension of the interface between aluminum and graphite, reported in our previous paper \cite{reshetniak2020aluminumcarbon}.

Since the aluminum melt wets the fullerene surface poorly, the molecules distributed in the melt tend to form bound clusters. We observed the coagulation in the direct MD simulations, carried out at time intervals of 10~ns. The estimated characteristic time of the coagulation was $t_{1/2} \approx 18.0$~ns. It is much longer than the relaxation time of the diffusion distribution of fullerenes by volume, $t_{D} \approx 0.29$~ns. Thus it is possible to separate the ``fast'' process of the diffusion distribution and the ``slow'' one of the coagulation for approximate analysis of a metastable system using methods of equilibrium statistical physics.

The RDF of the centers of mass of fullerenes has pronounced maxima and minima, which indicates the strong coupling between the molecules. The interaction has a capillary nature and is performed through the elastic stress fields in the aluminum melt.

Analysis of the coagulation kinetics confirms the strong capillary interaction between the fullerenes. When applying the Smoluchowski theory, the coagulation rate constant appears to be overestimated by about 100 times compared to the results of the MD simulations. Consequently, the coagulation process has an activation barrier, which occurs due to a long-range intermolecular interaction potential.

We calculated the values of the CTE and the bulk modulus of fullerenes. In the temperature range from 100 to 2000~K we obtained the constant value of the CTE, $\beta = 1.15 \times 10^{-5}$~K$^{-1}$. This result differs from the data of Ref~\cite{kwon_thermal_2004}, where the authors reported a strong nonlinear dependence $\beta (T)$ for $T \le 300$~K, and at $T \geq 300$~K $\beta$ slowly changed from $1.8 \times 10^{-5}$ to $2.0 \times 10^{-5}$~K$^{-1}$. The discrepancies obtained at $100 \le T \le 300$~K are caused by applying the Tersoff potential in this work, which does not reproduce the nonlinear temperature dependence of $\beta$. The origin of the discrepancies at $T \geq 300$~K is the choice of the method for calculating the fullerene volume. We estimated the volume using the radius of the interface $R_{f}$, considering the finite atomic radii. In contrast, in \cite{kwon_thermal_2004} the authors used the average distance $R$ from the fullerene center to the carbon atoms. Recalculation of the CTE using $R$ as the fullerene radius yields a value of $\beta = 1.63 \times 10^{-5}$~K$^{-1}$, reasonably coinciding with the data from Ref~\cite{kwon_thermal_2004}. 

The calculated value of the bulk modulus of fullerene is $K_{T} = 162$~GPa. This value differs markedly from the data available in the literature \cite{khabibrakhmanov_2020,peon2014,kaur2010,ruoff1991bulk}, which is only partly explained by the dependence of the bulk modulus on the choice of the method for calculating the nanoparticle volume. Recalculation using the average distance from carbon atoms to the center of mass of the molecule as the fullerene radius yields 324~GPa. An analysis of the literature data \cite{kaur2010} also testifies to the sensitivity of the results of calculating the elastic moduli to the choice of the interaction model for carbon atoms.

\begin{acknowledgments}
This work was supported by The Ministry of Science and Higher Education of the Russian Federation (Agreement with Joint Institute for High Temperatures RAS No 075-15-2020-785).
\end{acknowledgments}


\nocite{*}

\bibliography{apssamp}

\end{document}